\documentclass[nohyper]{JHEP3}
\usepackage{cite,epsfig,amssymb,amsmath}

\newcommand{\mycomm}[1]{\hfill\break $\phantom{a}$\kern-3.5em{\tt===$>$ \bf #1}\hfill\break}
\newcommand{\mycommA}[1]{\hfill\break $\phantom{a}$\kern-3.5em{\tt   $>$ \bf #1}\hfill\break}

\newcommand{\be}{\begin{equation}}
\newcommand{\ee}{\end{equation}}
\newcommand{\ba}{\begin{eqnarray}}
\newcommand{\ea}{\end{eqnarray}}

\catcode`\@=11 % This allows us to modify PLAIN macros.
\def\lsim{\mathrel{\mathpalette\@versim<}}
\def\gsim{\mathrel{\mathpalette\@versim>}}
\def\@versim#1#2{\vcenter{\offinterlineskip
        \ialign{$\m@th#1\hfil##\hfil$\crcr#2\crcr\sim\crcr } }}
\catcode`\@=12 % at signs are no longer letters

\title{On threshold resummation of singlet structure and fragmentation  functions}

\author{G. Grunberg\\
        Centre de Physique Th\'eorique,  \'Ecole
Polytechnique, CNRS,\\
        91128 Palaiseau Cedex, France\\
        E-mail: \email{georges.grunberg@cpht.polytechnique.fr}}

\abstract{The large-$x$ behavior of the physical evolution kernels appearing in the second order evolution equations of the singlet  $F_2$  structure function and of the $F_{\phi}$ structure function in $\phi$-exchange DIS is investigated. The validity of a leading logarithmic threshold resummation,  analogous to the one prevailing  for the non-singlet physical kernels, is established, allowing to recover  the predictions of Soar et al. for the double-logarithmic  contributions ($\ln^i(1-x)$, $i=4,5,6$) to the four loop  splitting functions $P^{(3)}_{qg}(x)$ and $P^{(3)}_{gq}(x)$. Threshold resummation  at the   next-to-leading logarithmic level is found however to  break down in the three loop kernels, except in the ``supersymmetric'' case $C_A=C_F$. Assuming a full threshold resummation does hold in this case also beyond three loop gives some information   on the leading and next-to-leading single-logarithmic contributions ($\ln^i(1-x)$, $i=2,3$) to $P^{(3)}_{qg}(x)$ and $P^{(3)}_{gq}(x)$.  Similar results are obtained for singlet fragmentation functions in $e^+e^-$ annihilation up to two loop, where a large-$x$ Gribov-Lipatov relation in  the physical kernels  is  pointed out. Assuming this relation  also holds at three loop, one gets predictions for all large-$x$ logarithmic contributions to the three loop timelike splitting function $P^{(2)T}_{gq}(x)$, which are related to similar terms in $P^{(2)}_{qg}(x)$.}

\preprint{ }

\begin{document}

\section{Introduction}
Although threshold resummation of  non-singlet structure and fragmentation functions is by now  well established \cite{Sterman:1986aj,Catani:1989ne,Cacciari:2001cw,Moch:2009my} in QCD, there is still not a comparable understanding of the similar problem in the singlet case. Motivated by some recent progress   \cite{Soar:2009yh,Vogt:2010cv,Almasy:2010wn}, this issue is addressed anew in the present paper. As in previous studies \cite{vanNeerven:2001pe,Gardi:2007ma,Grunberg:2007nc,Moch:2009mu,Grunberg:2009yi,Moch:2009hr,Grunberg:2009am,Grunberg:2009vs,Grunberg:2010sw} of the non-singlet case,  the present approach, as well as that of \cite{Soar:2009yh}, focuses on the properties of ``physical evolution kernels'' \cite{Furmanski:1981cw, Catani:1996sc,Blumlein:2000wh}. At the difference of \cite{Soar:2009yh} however, which uses a matrix   kernel  requiring the simultaneous consideration of another process (scalar $\phi$-exchange Deep Inelastic Scattering (DIS)) along with photon-exchange DIS, the present work adopts a more ``intrinsic'' point of view which decouples the two processes: namely, it deals with the study of the  scalar physical evolution kernels (defined in section 2 and 5)  occurring in the second order evolution equations \cite{Furmanski:1981cw,Blumlein:2000wh}  associated respectively to the singlet $F_2$  and $F_{\phi}$ structure functions. The large-$N$ behavior of the  $F_2$ kernels is investigated in section 3, where it is  pointed out that at large $N$  there is actually only {\em one} independent singlet  physical  kernel. In section 4, an obstruction to a standard form \cite{Gardi:2007ma}  of threshold resummation in momentum space is observed in the three loop $F_2$ kernel at the next-to-leading logarithmic level, which is found to be removed in the ``supersymmetric'' case $C_A=C_F$.
Assuming threshold resummation at $C_A=C_F$ does hold beyond three loop, the ensuing large-$N$ predictions for the four loop off-diagonal anomalous dimension $\gamma_{qg}^{(3)}(N)$ are derived. The corresponding  predictions of  \cite{Soar:2009yh} for the double logarithmic terms (now confirmed in \cite{Vogt:2010cv,Almasy:2010wn})  are recovered, and some additional  information is obtained concerning the   single logarithmic terms. A similar study of the $F_{\phi}$ kernels is performed in section 5, yielding analogous large-$N$ information on  $\gamma_{gq}^{(3)}(N)$.
 Fragmentation functions in $e^+e^-$ annihilation  are dealt with in section 6, where a large-$x$ Gribov-Lipatov relation between the spacelike and  timelike physical kernels is pointed out at two loop. Assuming  a similar relation is valid at three loop, {\em all} large-$x$ logarithmic contributions to the three loop timelike splitting function $P^{(2)T}_{gq}(x)$ are predicted. The conclusions, and some additional discussion, are presented in section 7. The connection between the present approach and that of \cite{Soar:2009yh} is explained in Appendix A.

\section{Physical evolution kernels of the $F_2$ singlet structure function}

Consider the  singlet $F_2$ structure function in Mellin moment space:

\begin{equation}F(N,Q^2)=\int_0^1dx \ x^{N-1}F_2(x,Q^2)/x\label{F}\ .\end{equation}
At large $Q^2$, it satisfies the standard OPE representation:

\begin{equation} F(N,Q^2)= <e^2_q> \Big(C_q(N,Q^2/\mu^2,a_s) q(N,\mu^2)+C_g(N,Q^2/\mu^2,a_s) g(N,\mu^2)\Big)\label{OPE}\ ,\end{equation}
where $C_a(N,Q^2/\mu^2,a_s)$  ($a=q,g$) are the singlet quark and gluon coefficient functions, $q(N,\mu^2)=\sum_{i=1}^{n_f}(q_i+\bar{q}_i)$ the singlet quark distribution, $g(N,\mu^2)$ the gluon   distribution, and  the factorization scale in the coefficient functions has been chosen to be equal to the renormalization scale $\mu^2$ with $a_s= \alpha_s(\mu^2)/4\pi$. Taking two derivatives of eq.(\ref{OPE}) with respect to $Q^2$, and eliminating the quark and gluon distributions, one obtains \cite{Furmanski:1981cw,Blumlein:2000wh} the second order physical evolution equation:

\begin{equation}\ddot {F}(N,Q^2)=K(N,Q^2)\dot {F}(N,Q^2)+J(N,Q^2)F(N,Q^2)\ ,\label{physkernels}
\end{equation}
where  $\dot {F}\equiv \partial F/\partial\ln Q^2$, which defines the singlet  physical evolution kernels $K(N,Q^2)$ and $J(N,Q^2)$.
$K$ and $J$ are renormalization group (and scheme) invariant quantities, which are obtained as combinations of   coefficient functions. One gets:

\begin{equation}K(N,Q^2)=\frac{C_q(N,Q^2/\mu^2,a_s)\ddot {C}_g(N,Q^2/\mu^2,a_s)-C_g(N,Q^2/\mu^2,a_s)\ddot {C}_q(N,Q^2/\mu^2,a_s)}{C_q(N,Q^2/\mu^2,a_s)\dot {C}_g(N,Q^2/\mu^2,a_s)-C_g(N,Q^2/\mu^2,a_s)\dot {C}_q(N,Q^2/\mu^2,a_s)}\label{physkernel-K}\ ,
\end{equation}
and

\begin{equation}J(N,Q^2)=-\frac{\dot{C}_q(N,Q^2/\mu^2,a_s)\ddot {C}_g(N,Q^2/\mu^2,a_s)-\dot{C}_g(N,Q^2/\mu^2,a_s)\ddot {C}_q(N,Q^2/\mu^2,a_s)}{C_q(N,Q^2/\mu^2,a_s)\dot {C}_g(N,Q^2/\mu^2,a_s)-C_g(N,Q^2/\mu^2,a_s)\dot {C}_q(N,Q^2/\mu^2,a_s)}\label{physkernel-J}\ ,
\end{equation}
where $\dot {C_a}\equiv \partial C_a/\partial\ln Q^2$ (at {\em fixed} $\mu^2\neq Q^2$). Moreover, the renormalization group equations for the coefficient functions yield:

\begin{equation}\dot {C}_q(N,Q^2/\mu^2,a_s)=\beta(a_s)\frac{\partial C_q(N,Q^2/\mu^2,a_s)}{\partial a_s}-\gamma_{qq}(N,a_s)C_q(N,Q^2/\mu^2,a_s)-\gamma_{gq}(N,a_s)C_g(N,Q^2/\mu^2,a_s)\ ,\label{dCq}\end{equation}

\begin{equation}\dot {C}_g(N,Q^2/\mu^2,a_s)=\beta(a_s)\frac{\partial C_g(N,Q^2/\mu^2,a_s)}{\partial a_s}-\gamma_{qg}(N,a_s)C_q(N,Q^2/\mu^2,a_s)-\gamma_{gg}(N,a_s)C_g(N,Q^2/\mu^2,a_s)\ ,\label{dCg}
\end{equation}
where the moment space anomalous dimensions are related to the splitting functions by the standard convention:

\begin{equation}\gamma_{ab}(N,a_s)=-\int_0^1 dx \ x^{N-1} P_{ab}(x,a_s)\label{convention}\ , \end{equation}
so that

\begin{equation} \Big( \begin{array}{c} \!\! \dot{q}(N,\mu^2) \!\! \\ \! \dot{g}(N,\mu^2) \!\!
        \end{array} \Big) \:=\:
- \Big( \! \begin{array}{cc}\gamma_{qq}(N,a_s)\! & \gamma_{qg}(N,a_s)\! \\
  \gamma_{gq}(N,a_s)\! & \gamma_{gg}(N,a_s)\! \end{array}\! \Big) \Big( \begin{array}{c} \!\! q(N,\mu^2)\!\! \\ \! g(N,\mu^2) \!\!  \end{array} \Big)
\label{A-P}\ .  \end{equation}
Taking one $Q^2$ derivative of eqs.(\ref{dCq}) and (\ref{dCg}) one also gets:
\begin{equation}\ddot {C}_q(N,Q^2/\mu^2,a_s)=\beta(a_s)\frac{\partial \dot{C}_q(N,Q^2/\mu^2,a_s)}{\partial a_s}-\gamma_{qq}(N,a_s)\dot{C}_q(N,Q^2/\mu^2,a_s)-\gamma_{gq}(N,a_s)\dot{C}_g(N,Q^2/\mu^2,a_s)\ ,\label{ddCq}\end{equation}
and:

\begin{equation}\ddot {C}_g(N,Q^2/\mu^2,a_s)=\beta(a_s)\frac{\partial \dot{C}_g(N,Q^2/\mu^2,a_s)}{\partial a_s}-\gamma_{qg}(N,a_s)\dot{C}_q(N,Q^2/\mu^2,a_s)-\gamma_{gg}(N,a_s)\dot{C}_g(N,Q^2/\mu^2,a_s)\ .\label{ddCg}
\end{equation}
Inserting eqs.((\ref{dCq})-(\ref{dCg})) and ((\ref{ddCq})-(\ref{ddCg})) into eqs.(\ref{physkernel-K}) and (\ref{physkernel-J}), then setting $\mu^2=Q^2$ (which is legitimate since $K$ and $J$ are renormalization group invariant quantities), and using the expansions  of the quark and gluon coefficient functions:

\begin{equation}C_q(N,1,a_s)=1+\sum_{i=1}^{\infty}c_q^{(i)}(N)a_s^i\label{Cq-expand}\end{equation}

\begin{equation}C_g(N,1,a_s)=\sum_{i=1}^{\infty}c_g^{(i)}(N)a_s^i\label{Cg-expand}\ ,\end{equation}
as well as those of the beta function and of the anomalous dimensions:

\begin{equation}\beta(a_s)=-\sum_{i=0}^{\infty}\beta_i a_s^{i+2}\label{beta-expand}\end{equation}

\begin{equation}\gamma_{ab}(N,a_s)=\sum_{i=0}^{\infty}\gamma_{ab}^{(i)}(N)a_s^{i+1}\label{gamma-expand}\end{equation}
with $(a,b)=(q,g)$,
one  obtains:

\begin{equation}K(N,Q^2)=\sum_{i=0}^{\infty}K^{(i)}(N) a_s^{i+1}
\label{K-expand}\end{equation}
and

\begin{equation}J(N,Q^2)=\sum_{i=0}^{\infty}J^{(i)}(N) a_s^{i+2}
\label{J-expand}\ ,\end{equation}
where $a_s=a_s(Q^2)$, and $K^{(i)}(N)$ and $J^{(i)}(N)$ are expressed as (rather long) combinations of the coefficient functions and anomalous dimensions expansion coefficients.

\section{Large-N}
Let us now consider the large-$N$ limit. Using the known  large-$N$  expansions of the anomalous dimensions and of the coefficient functions, one deduces the  large-$N$  expansions of $K^{(i)}(N)$ and $J^{(i)}(N)$. In the following, I shall focus on $K^{(i)}(N)$  only, since one can show (Appendix A.1) that at large  $N$, $J(N,Q^2)$ can be expressed in term of   $K(N,Q^2)$ and   the  physical evolution kernel $K_{ns}(N,Q^2)$ of the non-singlet $F_{2}$ structure function $F_{ns}$:

\begin{equation}J(N,Q^2)\sim {\dot K}_{ns}(N,Q^2)-K_{ns}(N,Q^2)\Big[K(N,Q^2)-K_{ns}(N,Q^2)\Big]+{\cal O}(1/N^2)\label{J-largeN}\ ,
\end{equation}
with
\begin{equation}\dot{F}_{ns}(N,Q^2)=K_{ns}(N,Q^2)F_{ns}(N,Q^2)\ .\label{ns-physkernel}
\end{equation} 
Thus at  {\em large} $N$, only $K^{(i)}(N)$ and  $K_{ns}^{(i)}(N)$ (with $K_{ns}(N,Q^2)=\sum_{i=0}^{\infty}K_{ns}^{(i)}(N) a_s^{i+1}$) are independent functions. In particular, eq.(\ref{physkernels}) can be solved at large $N$ as:

\begin{equation}\frac{\partial (\dot {F}-K_{ns}F)}{\partial\ln Q^2}\sim K_{ps}(N,Q^2)(\dot {F}-K_{ns}F)\ ,\label{large-N-sol}
\end{equation}
where $K_{ns} F=\frac{\dot{F}_{ns}}{F_{ns}}F$  from eq.(\ref{ns-physkernel}), and I defined:

\begin{equation}K_{ps}(N,Q^2)\equiv K(N,Q^2)-K_{ns}(N,Q^2)\label{K-ps}\ .
\end{equation}
$K_{ps}$ can be  viewed as a ``pure singlet'' contribution to $K$.
Since threshold resummation \cite{Sterman:1986aj,Catani:1989ne}  is well established \cite{Gardi:2007ma,Friot:2007fd} for   $K_{ns}(N,Q^2)$, one can consider only\footnote{It should be noted that the large $N$  singlet quark coefficients (eq.(\ref {iloop-cq}) below) coincide \cite{Vermaseren:2005qc} with the non-singlet ones, thus threshold resummation of $K_{ns}(N,Q^2)$  is related to threshold resummation of the singlet quark coefficient function itself.}    $K(N,Q^2)$.

a)\ \underbar{One loop}:

\noindent For the one loop physical kernel coefficient $K^{(0)}(N)$ one finds:

\begin{equation}K^{(0)}(N)=-\gamma_{qq}^{(0)}(N)-\gamma_{gg}^{(0)}(N)-\beta_0\label{K0}\ ,
\end{equation}
which depends  on the one loop {\em diagonal} anomalous dimensions (and $\beta_0=\frac{11}{3}C_A-\frac{2}{3}n_f$).
Using the large-$N$ asymptotics \cite{Korchemsky:1988si}:

\begin{equation}\gamma_{aa}^{(i)}(N)\sim A_{i+1}^a\ln \bar{N}-B_{i+1}^a\label{iloop-aa}
\end{equation}
($a=q,g$) where $\ln \bar{N}=\ln N+\gamma_E$, one deduces:

 \begin{equation}K^{(0)}(N)\sim K_{11}\ln \bar{N}+K_{10}\label{K0-as}\ ,
\end{equation}
with:

 \begin{eqnarray}K_{11}&=& -(A_1^g+A_1^q)\label{K11}\\
 K_{10}&=&B_1^g+B_1^q-\beta_0 \label{K10}\ ,
\end{eqnarray}
where \cite{Altarelli:1977zs} $A_1^q=4C_F$, $A_1^g=4C_A$, $B_1^q=3C_F$, $B_1^g=\beta_0$.
One thus gets:

\begin{eqnarray}K_{11}&=& -4(C_A+C_F)\label{K11-value}\\
K_{10}&=&3C_F \label{K10-value}\ .
\end{eqnarray}

b)\ \underbar{Two loop}:

\noindent 
Beyond one loop, the $K^{(i)}(N)$ 's depend also on the off-diagonal anomalous dimensions, as well as on the quark and gluon coefficient functions.
Using eq.(\ref{iloop-aa}) and the large-$N$ asymptotics of the off-diagonal anomalous dimensions \cite{Vogt:2004mw}:

\begin{eqnarray}\gamma_{qg}^{(i)}(N)&\sim&\frac{1}{N}\sum_{j=0}^{2i} D_{i+1 j}\ln^j \bar{N}\label{iloop-qg}\\
\gamma_{gq}^{(i)}(N)&\sim&\frac{1}{N} \sum_{j=0}^{2i} \Delta_{i+1 j}\ln^j \bar{N}\label{iloop-gq}
\end{eqnarray}
(with $D_{10}=-2 n_f$, $\Delta_{10}=-2 C_F$), and of the coefficient functions \cite{Vermaseren:2005qc}:

\begin{eqnarray}c_q^{(i)}(N)&\sim&\sum_{j=0}^{2i}c_{qj}^{(i)}\ln^j \bar{N} \label{iloop-cq}\\
c_g^{(i)}(N)&\sim&\frac{1}{N}\sum_{j=0}^{2i-1}c_{gj}^{(i)}\ln^j \bar{N} \label{iloop-cg}\ ,\end{eqnarray}
one gets for $i\geq 1$:

 \begin{equation}K^{(i)}(N)\sim\sum_{j=0}^{2i} K_{i+1 j}\ln^j \bar{N}\label{Ki-as}\ .
\end{equation}
At two loop ($i=1$) one finds:

 \begin{eqnarray}K_{22}&=&\frac{1}{2}\tilde{D}_{22}\beta_0-\Big[2 c_{q2}^{(1)}-\frac{1}{2}\tilde{c}_{g1}^{(1)}(A_1^g-A_1^q)\Big]\beta_0\label{K22}\\
 K_{21}&=&\frac{1}{2}\tilde{D}_{21}\beta_0-(A_2^g+A_2^q)\label{K21}\\
 & &-\Big[2 c_{q1}^{(1)}-\frac{1}{2}\tilde{c}_{g0}^{(1)}(A_1^g-A_1^q)+\frac{1}{2}\tilde{c}_{g1}^{(1)}(B_1^g-B_1^q-\beta_0)\Big]\beta_0\nonumber\\
  K_{20}&=&\frac{1}{2}\tilde{D}_{20}\beta_0+(B_2^g+B_2^q)-\beta_1\label{K20}\\
 & &-\Big[2 c_{q0}^{(1)}+\frac{1}{2}\tilde{c}_{g0}^{(1)}(B_1^g-B_1^q-\beta_0)\Big]\beta_0\nonumber\ ,
 \end{eqnarray}
 with:
 
  \begin{eqnarray}\tilde{D}_{ji}&= &D_{ji}/n_f=-2D_{ji}/D_{10}\label{norm-D}\\
  \tilde{c}_{gi}^{(j)}&=&c_{gi}^{(j)}/n_f=-2c_{gi}^{(j)}/D_{10}\label{norm-cg}\ ,\end{eqnarray}   
  and the brackets contain only one loop quantities. Using   known results on one loop coefficient functions and two loop anomalous dimensions  (see e.g.   \cite{Moch:2004pa,Vogt:2004mw,Vermaseren:2005qc} and references therein), one gets:

\begin{eqnarray}K_{22}&=& -2(C_A+C_F)\beta_0\label{K22-value}\\
K_{21}&=&\Big(-\frac{16}{3}+8\zeta_2\Big)C_A^2 +\Big(-\frac{16}{3}+8\zeta_2\Big)C_A C_F -\frac{32}{3}C_A\beta_0-\frac{35}{3}C_F\beta_0\label{K21-value}\\
K_{20}&=&(-4+12\zeta_3)C_A^2 +(-10-12\zeta_3)C_A C_F+\Big(\frac{3}{2}-12\zeta_2+24\zeta_3\Big)C_F^2 \nonumber\\
& &+2 C_A\beta_0+\Big(\frac{27}{2}+10\zeta_2\Big)C_F\beta_0-\beta_1 \label{K20-value}\ ,
\end{eqnarray}
with $\beta_1=-7 C_A^2-11 C_A C_F+(5 C_A+3 C_F)\beta_0$.

c)\ \underbar{Three loop}:

\noindent 
At three loop, one finds:

 \begin{eqnarray}K_{3i}&=&\tilde{D}_{3i}\beta_0+(lower\ loop\  quantities)\ (2\leq i\leq 4)\label{K3i}\\
 K_{31}&=&\tilde{D}_{31}\beta_0-(A_3^g+A_3^q)+(lower\ loop\  quantities) \label{K31}\\
 K_{30}&=&\tilde{D}_{30}\beta_0+(B_3^g+B_3^q)-\beta_2+(lower\ loop\  quantities) \label{K30}\ .
 \end{eqnarray}
Using results on two loop coefficient functions (see e.g. \cite{Moch:1999eb,Vermaseren:2005qc,Soar:2009yh} and references therein) and three loop anomalous dimensions \cite{Moch:2004pa,Vogt:2004mw}, one gets:

\begin{eqnarray}K_{34}&=& 0\label{K34-value}\\
K_{33}&=& -\frac{4}{3}(C_A+C_F)\beta_0^2\label{K33-value}\\
K_{32}&=&\Big(-\frac{112}{3}+24\zeta_2\Big)C_A^2\beta_0 +\Big(\frac{128}{3}-8\zeta_2\Big)C_A C_F\beta_0 -16 C_F^2\beta_0-\frac{32}{3}C_A\beta_0^2-\frac{35}{3}C_F\beta_0^2\nonumber\\
& &-2( C_A+ C_F)\beta_1\label{K32-value}\\
K_{31}&=&\Big(-\frac{856}{9}+\frac{128}{3}\zeta_2-24\zeta_3\Big)C_A^2\beta_0 +\Big(-\frac{1393}{9}+\frac{32}{3}\zeta_2+168\zeta_3\Big)C_A C_F\beta_0 +(-64+72\zeta_2-96\zeta_3) C_F^2\beta_0\nonumber\\
& &+\Big(-\frac{238}{9}+8\zeta_2\Big)C_A\beta_0^2+\Big(-\frac{313}{9}+4\zeta_2\Big)C_F\beta_0^2-4 C_A\beta_1-5 C_F\beta_1-2\beta_0\beta_1\label{K31-value}\\
& &+\Big(\frac{224}{9}+\frac{64}{3}\zeta_2-\frac{176}{5}\zeta_2^2+176\zeta_3\Big)C_A^3+\Big(\frac{2039}{9}+\frac{64}{3}\zeta_2-\frac{176}{5}\zeta_2^2\Big)C_A^2 C_F+\Big(\frac{605}{3}-176\zeta_3\Big)C_A  C_F^2\nonumber\ .\end{eqnarray}

d)\ \underbar{Four loop}:

\noindent 
For future use, I also quote the large-$N$ structures at the four loop level.
One finds:

 \begin{equation}K_{4i}= \frac{3}{2}\tilde{D}_{4i}\beta_0+(lower\ loop\  quantities)\ (2\leq i\leq 6)\label{K4i}
 \ , \end{equation}
 where I restricted to the range $2\leq i\leq 6$ which will be of interest in the following.
The ``lower loop quantities'' in  eqs.(\ref{K3i}), (\ref{K31}), (\ref{K30}) and (\ref{K4i}) are given by very long expressions, which need not be written down explicitly here, and depend {\em both} on anomalous dimensions and (quark and gluon) coefficient functions. One should mention however that these  expressions do not involve the $\Delta_{ij}$'s coefficients, therefore the $K_{ij}$'s (hence the $K^{(i)}(N)$'s at {\em large} $N$) are {\em independent} of the off-diagonal $\gamma_{gq}(N)$ anomalous dimension\footnote{This is not the case for the $K^{(i)}(N)$'s  at {\em finite} $N$ for $i\geq 2$.}. Moreover, the dependence upon the one loop coefficient $D_{10}=-2 n_f$ reveals itself only through the appearance of the ``renormalized'' off-diagonal coefficients $\tilde{D}_{ij}$ and $\tilde{c}_{gj}^{(i)}$ (eqs.(\ref{norm-D}) and (\ref{norm-cg})).

\section{Momentum space results and threshold resummation}

\subsection{An obstruction to threshold resummation}
In this section, a few observations are presented.

1) The momentum space  kernel $K(x, Q^2)$ is related in the standard way to the Mellin moment space one by:

\begin{equation} K(N,Q^2)\equiv\int_0^1dx\, x^{N-1} K(x, Q^2) \label{K-tilde1}\ .\end{equation} 
Usually, the moment space kernel $K(N,Q^2)$  cannot  be inverted analytically to momentum space (due to the presence of $N$-dependent functions in the denominators of eq.(\ref{physkernel-K})). However, this is no more the case at large $N$, as is clear from the expressions in the previous section.

2) Using the large-$x$-large-$N$ correspondence:

 \begin{equation}\frac{\ln^n(1-x)}{(1-x)_+} <->\frac{(-1)^{n+1}}{n+1}\ln^{n+1}N+...\label{x-N}\ ,\end{equation}
the momentum space  physical kernel expansion coefficients $K^{(i)}(x)$  (obtained by inversion of the moment space ones) are seen for $0\leq i\leq 2$  to have only a single-logarithmic enhancement at large-$x$ (cf. in particular eq.(\ref{K34-value})), similarly to the  non-singlet kernel \cite{Grunberg:2009yi,Moch:2009hr}, namely:

\begin{equation}K^{(i)}(x)={\cal O}\Big (\frac{\ln^i(1-x)}{(1-x)_+}\Big )\label{Ki-large-x}
\end{equation}

3) Consequently, it makes sense to ask whether an $x$ space threshold resummation, similar to the one valid \cite{Gardi:2007ma,Grunberg:2009yi} for the non-singlet physical kernel $K_{ns}(x,Q^2)$, is also possible here, namely whether for $x\rightarrow 1$, $(1-x) K(x,Q^2)$ effectively depends only on the {\em single} scale $(1-x)Q^2$ :

\begin{equation}
\label{K-conjecture}
K(x,Q^2)\sim \frac{{\cal J}\left((1-x)Q^2\right)}{(1-x)_+}\ , \end{equation}
where ${\cal J}\left(Q^2\right)$ is a renormalization group invariant quantity, so that:

\begin{eqnarray}{\cal J}\left((1-x)Q^2\right)&=&j_{1}\ a_s+a_s^2[-j_{1}\beta_0 L_x+j_{2}]\label{j-expand}\\
&+ &a_s^3[j_{1}\beta_0^2 L_x^2-(j_{1}\beta_1+2 j_{2} \beta_0) L_x+j_{3}]\nonumber\\
&+ &a_s^4[-j_{1}\beta_0^3 L_x^3+(\frac{5}{2}j_{1}\beta_1\beta_0+3 j_{2} \beta_0^2) L_x^2-(j_{1}\beta_2+2 j_{2} \beta_1+3 j_{3} \beta_0) L_x+j_{4}]+...\ ,\nonumber
\end{eqnarray}
where $a_s=a_s(Q^2)$ and $L_x\equiv \ln(1-x)$.
Comparing eqs.(\ref{K-conjecture}) and (\ref{j-expand}) with the exact result (known up to ${\cal O}(a_s^3)$), inverting to momentum space the results of section 3 to get the leading ${\cal O}(1/(1-x))$ term in $K(x,Q^2)$, one finds :

i) The leading logarithms in the  exact result agree with the leading logarithms in  eq.(\ref{j-expand}), with

\begin{equation}j_{1}=-K_{11}=4(C_A+C_F)\label{j1}\ .\end{equation}

ii) There is a discrepancy at the next-to-leading logarithmic  level, starting at  ${\cal O}(a_s^3)$, namely:

\noindent Comparing the ${\cal O}(a_s^2)$ term in eq.(\ref{j-expand}) with the corresponding term in the exact result, one determines

\begin{equation}j_{2}=-K_{21}=(\frac{16}{3}-8\zeta_2)(C_A + C_F)C_A +\frac{32}{3}C_A\beta_0+\frac{35}{3}C_F\beta_0\label{j2}\  .\end{equation}
However, comparing with the exact coefficient of the  single $\frac{\ln(1-x)}{1-x}$ term occurring at ${\cal O}(a_s^3)$ in $K(x,Q^2)$, one finds the latter is {\em not} given by $-(j_{1}\beta_1+2 j_{2} \beta_0)$ as required by eq.(\ref{K-conjecture}) and (\ref{j-expand}), but rather by:

\begin{equation}2 K_{32}=-[j_{1}\beta_1+2 (j_{2}+\delta) \beta_0]\label{discrepancy}\ ,\end{equation}
with

\begin{equation}\delta=16(C_A-C_F)[(2-\zeta_2)C_A-C_F]\label{delta2}\ .\end{equation}
This discrepancy represents an obstruction to a standard threshold resummation at the next-to-leading logarithmic level of the singlet physical kernel. The situation here looks similar to the one occuring \cite{Moch:2009mu,Grunberg:2009am} in the case of the (non-singlet) $F_L$ structure function \cite{Akhoury:1998gs,Akhoury:2003fw}.

\subsection{A conjectured threshold resummation for $C_A=C_F$ and four loop predictions}
It is remarkable that the  obstruction $\delta$ to threshold resummation at the next-to-leading logarithmic level {\em vanishes}  in the ``supersymmetric'' case $C_A=C_F$. This fact suggests that a standard  threshold resummation, similar to the one available for the non-singlet physical kernel  $K_{ns}(x,Q^2)$, might indeed also be possible for the singlet kernel in this case. 

Let us  derive the  four loop predictions following by assuming the validity of the ansatz (\ref{K-conjecture}) and (\ref{j-expand}). As a warm-up, I consider first the two and three loop predictions. Eqs.(\ref{K-conjecture}) and (\ref{j-expand}) imply, after converting to Mellin moment space:

\noindent a)\ \underbar {Two  loop prediction}:
$K_{22}=\frac{1}{2}K_{11}\beta_0$, from the ${\cal O}(a_s^2 L_x)$ leading  logarithm (LL) term in eq.(\ref{j-expand}), in agreement  with the exact result (eq.(\ref{K22-value})), as expected from the observations in section 4.1. Using eq.(\ref{K22}), this constraint determines the two loop off-diagonal anomalous dimension coefficient $D_{22}$ from one loop quantities. One gets:

\begin{equation}D_{22}=4n_f C_{AF}\label{D22predict}\ ,\end{equation}
where $C_{AF}=C_A-C_F$,
which  agrees  as expected  (after straightforward inversion to $x$-space, eq.(\ref {convention})) with the exact result  \cite{Vogt:2004mw} for the double logarithmic term in $P_{qg}^{(1)}(x)$. 

\noindent b)\ \underbar {Three loop predictions}: 

i) $K_{34}=0$, from the absence of an ${\cal O}(a_s^3 L_x^3)$ double logarithmic term in eq.(\ref{j-expand}). Using eq.(\ref{K3i}),  this constraint determines $D_{34}$ from lower loop quantities. One gets:

\begin{equation}D_{34}=-\frac{4}{3}n_f C_{AF}^2\label{D34predict}\ ,\end{equation}
 which indeed agrees as expected  with the exact result \cite{Vogt:2004mw}.

ii) $K_{33}=\frac{1}{3}K_{11}\beta_0^2$, from the ${\cal O}(a_s^3 L_x^2)$ LL  term in eq.(\ref{j-expand}). Then eq.(\ref{K3i}) yields:

\begin{equation}D_{33}=\frac{2}{3}n_f C_{AF}(3 C_F-\beta_0)\label{D33predict}\ ,\end{equation}
which again agrees  as expected with the exact result  \cite{Vogt:2004mw}. Thus double logarithmic terms ($\ln^i(1-x)$, $i=3,4$) in the off-diagonal three loop splitting function $P_{qg}^{(2)}(x)$ are correctly reproduced.

iii) $K_{32}=\frac{1}{2}(K_{11}\beta_1+2 K_{21}\beta_0)$, from the ${\cal O}(a_s^3 L_x)$ next-to-leading (NLL) logarithm in eq.(\ref{j-expand}). Eq.(\ref{K3i}) then yields:

\begin{equation}D_{32}\vert{predict}=\frac{4}{3}n_f C_{AF}[5\beta_0+(28-24\zeta_2)C_A+9C_F]\label{D32predict}\ .\end{equation}
This time, the above prediction disagrees for $C_A\neq C_F$ with the exact result  \cite{Vogt:2004mw}:

\begin{equation}D_{32}=\frac{4}{3}n_f C_{AF}[5\beta_0+(4-12\zeta_2)C_A+21C_F]\label{D32}\ ,\end{equation}
as expected from the fact that $\delta\neq 0$. I note however that for $C_A=C_F$, $D_{32}$  is correctly  (as again expected) predicted to vanish. Since $D_{34}$ and $D_{33}$ also vanish in this limit, it follows after inversion to $x$-space that the leading single logarithmic term ($\ln^2(1-x)$) in  $P_{qg}^{(2)}(x)$ also vanishes for  $C_A=C_F$.
This (a priori surprising) fact finds a nice interpretation in the present framework: it is seen to follow from the correctness (up to ${\cal O}(a_s^3)$) of   eq.(\ref{K-conjecture}) and (\ref{j-expand}) for $C_A=C_F$.

\noindent Moreover, the large $\beta_0$ term in eq.(\ref{D32predict}) is also correct. This result should be expected since the discrepancy  $\delta$ (eq.(\ref{delta2})) is non-leading at large $\beta_0$. Thus one might conjecture that the ansatz (\ref{K-conjecture}) and (\ref{j-expand}) is  valid at large $\beta_0$ even for $C_A\neq C_F$.

\noindent c)\ \underbar {Four loop predictions}: the ansatz (\ref{K-conjecture}) and (\ref{j-expand}) similarly implies at the ${\cal O}(a_s^4)$ level, after converting to Mellin moment space:

i) $K_{46}=K_{45}=0$, from the absence of  ${\cal O}(a_s^4 L_x^5)$ and ${\cal O}(a_s^4 L_x^4)$ double logarithmic terms in eq.(\ref{j-expand}). Using eq.(\ref{K4i}), these constraints determines the four loop off-diagonal anomalous dimension coefficients $D_{46}$ and $D_{45}$ from lower loop quantities. Using  known results for the three loop coefficient functions \cite{Soar:2009yh,Vermaseren:2005qc}, one gets:

\begin{eqnarray}D_{46}&=&0\label{D46predict}\\
D_{45}&=&n_f C_{AF}^2(-\frac{4}{3}C_F+\frac{2}{9}\beta_0)\label{D45predict}\ .\end{eqnarray}

ii) $K_{44}=\frac{1}{4}K_{11}\beta_0^3$, from the ${\cal O}(a_s^4 L_x^3)$ LL in eq.(\ref{j-expand}). Then eq.(\ref{K4i}) yields:

\begin{equation}D_{44}=n_f C_{AF}\Big[\Big(-\frac{40}{27}+\frac{80}{9}\zeta_2\Big)C_{AF}^2+\Big(-\frac{484}{27}+8\zeta_2\Big)C_{AF} C_F-\frac{80}{27}C_{AF}\beta_0+\frac{1}{2}C_F^2-\frac{1}{2}C_F\beta_0+\frac{1}{9}\beta_0^2\Big]
\label{D44predict}\ .\end{equation}
The predictions eqs.(\ref{D46predict})-(\ref{D44predict}), first obtained in \cite{Soar:2009yh}, have recently been shown to be correct in \cite{Vogt:2010cv,Almasy:2010wn}.
 Thus, after inversion to $x$-space, one finds that large $x$ double logarithmic terms ($\ln^i(1-x)$, $i=4,5,6$) in the off-diagonal $P_{qg}^{(3)}(x)$   splitting function are again correctly predicted by the present ansatz. The close  connection with the approach of \cite{Soar:2009yh} is explained in Appendix A.2.

iii) $K_{43}\vert{predict}=\frac{5}{6}K_{11}\beta_1\beta_0+ K_{21}\beta_0^2$, from the ${\cal O}(a_s^4 L_x^2)$ NLL  in eq.(\ref{j-expand}). Eq.(\ref{K4i}) then yields:

\begin{equation}D_{43}\vert{predict}=n_f C_{AF} \Big[-\frac{40}{27}\beta_0^2+\beta_0 P_1(C_A,C_F)+P_2(C_A,C_F)\Big]\label{D43predict}\ .\end{equation}
where $P_1$ and $P_2$ are respectively linear and quadratic homogeneous polynomials in $(C_A,C_F)$, which need not be written down since the corresponding terms are not expected to be correctly reproduced anyway by the present ansatz. However, the  vanishing for $C_A=C_F$ of  $D_{43}$ is again expected to be a valid prediction. Since $D_{45}$ and $D_{44}$ also vanish in this limit, the vanishing of $D_{43}$ implies, after inversion to $x$-space, the  vanishing for $C_A=C_F$  of the leading single logarithmic term ($\ln^3(1-x)$) in  $P_{qg}^{(3)}(x)$ (similarly to the three loop case). Moreover, the large $\beta_0$ term should also be correct.

iv) $K_{42}\vert{predict}=\frac{1}{2}K_{11}\beta_2+ K_{21}\beta_1+\frac{3}{2}K_{31}\beta_0$, from the ${\cal O}(a_s^4 L_x)$ NNLL in eq.(\ref{j-expand}). Eq.(\ref{K4i}) then yields:

\begin{eqnarray}D_{42}\vert{predict}&=&n_f\Big[\Big(\frac{40}{27}-\frac{2}{3}\zeta_2\Big)C_A\beta_0^2+\Big(-\frac{58}{27}-\frac{2}{3}\zeta_2\Big)C_F\beta_0^2+(26+2\zeta_2-16\zeta_3) C_F^2\beta_0\nonumber\\
& &+\Big(18-12\zeta_2-\frac{768}{5}\zeta_2^2-24\zeta_3+384\zeta_4\Big)C_F^3+C_{AF}\Big(\beta_0 Q_1(C_A,C_F)+Q_2(C_A,C_F)\Big)\Big]\nonumber\\
& &+\frac{2}{3}n_f \frac{\beta_1}{\beta_0}\delta\label{D42predict}\ ,\end{eqnarray}
where $Q_1$ and $Q_2$ are respectively linear and  quadratic  homogeneous polynomials in $(C_A,C_F)$ which again are not expected to be correctly predicted  by the present ansatz. The last term in eq.(\ref{D42predict}), which depends on $\delta$ (eq.(\ref{delta2})), is also obviously incorrect, and should be absent in the exact answer, since it is not even a  polynomial in the color factors. However, this term vanishes for $C_A=C_F$, which represents a {\em non-trivial consistency check} of the present proposal.

\noindent It should be noted that the single logarithm coefficients  $D_{21}$, $D_{31}$ and $D_{41}$ are {\em not} predicted in the present approach: rather, they  represent  input parameters for the threshold resummation, since they  determine respectively $K_{21}$, $K_{31}$ and $K_{41}$, hence the ${\cal O}(a_s^2)$, ${\cal O}(a_s^3)$ and ${\cal O}(a_s^4)$  constant terms $j_2$ (see  eq.(\ref{j2})),  $j_3$ and $j_4$ in eq.(\ref{j-expand}). It is also remarkable the {\em leading} single logarithmic coefficients $D_{32}$ and $D_{43}$ are both predicted to vanish when $C_A=C_F$, which may be the signal of a systematic structure.

\section{$\phi$-exchange DIS}

In the case of $\phi$-exchange DIS, quite similar formulas apply. Defining the moment space structure function by:

\begin{equation}F_{\phi}(N,Q^2)=\int_0^1dx \ x^{N-1}F_{\phi}(x,Q^2)\label{Fphi}\ ,\end{equation}
we have the large-$Q^2$  OPE representation \cite{Soar:2009yh}:

\begin{equation} F_{\phi}(N,Q^2)= C_{\phi,q}(N,Q^2/\mu^2,a_s) A_{q,nucl}(N,\mu^2)+C_{\phi,g}(N,Q^2/\mu^2,a_s) A_{g,nucl}(N,\mu^2)\label{phi-OPE}\ ,\end{equation}
where $C_{\phi,a}(N,Q^2/\mu^2,a_s)$  ($a=q,g$) are the  $\phi$-exchange  quark and gluon coefficient functions,  $A_{q,nucl}(N,\mu^2)$ and $A_{g,nucl}(N,\mu^2)$ the singlet quark and gluon  nucleon matrix elements, and $a_s= a_s(\mu^2)$. The physical evolutions kernels for this process are defined by:

\begin{equation}\ddot {F_{\phi}}(N,Q^2)=K_{\phi}(N,Q^2)\dot {F_{\phi}}(N,Q^2)+J_{\phi}(N,Q^2)F_{\phi}(N,Q^2)\ ,\label{phi-physkernels}
\end{equation}
and can be expanded as:

\begin{equation}K_{\phi}(N,Q^2)=\sum_{i=0}^{\infty}K_{\phi}^{(i)}(N) a_s^{i+1}
\label{Kphi-expand}\end{equation}
and

\begin{equation}J_{\phi}(N,Q^2)=\sum_{i=0}^{\infty}J_{\phi}^{(i)}(N) a_s^{i+2}
\label{Jphi-expand}\ ,\end{equation}
where $a_s= a_s(Q^2)$. As in the photon-exchange case, I shall concentrate in the following on $K_{\phi}(N,Q^2)$, as $J_{\phi}(N,Q^2)$ can be expressed at large $N$ in term of $K_{\phi}(N,Q^2)$ and a large-$N$ non-singlet ``gluonic'' kernel  $K_{\phi,ns}(N,Q^2)$ which satisfies a standard form of threshold resummation similarly to the non-singlet quark kernel $K_{ns}(N,Q^2)$ (see Appendix A.1).

\subsection{Large-$N$}
The main formal difference between the photon and the $\phi$-exchange cases is that the leading ${\cal O}(a_s^0)$ (resp. ${\cal O}(a_s)$) behavior of the quark (resp. gluon) coefficient functions (eqs.(\ref{Cq-expand}), (\ref{Cg-expand})) are interchanged, namely we have:

\begin{equation}C_{\phi,g}(N,1,a_s)=1+\sum_{i=1}^{\infty}c_{\phi g}^{(i)}(N)a_s^i\label{Cphig-expand}\end{equation}

\begin{equation}C_{\phi,q}(N,1,a_s)=\sum_{i=1}^{\infty}c_{\phi q}^{(i)}(N)a_s^i\label{Cphiq-expand}\ .\end{equation}
Consequently, at large $N$ the $K_{\phi}^{(i)}(N)$'s coefficients shall not depend  upon the off-diagonal $\gamma_{qg}(N)$ anomalous dimension, but rather on the $\gamma_{gq}(N)$ one. 

a)\ \underbar{One loop}:

\noindent 
At this order, the results are the same as in the photon-exchange case, namely:

 \begin{equation}K_{\phi}^{(0)}(N)\sim K_{11}^{\phi}\ln \bar{N}+K_{10}^{\phi}\label{Kphi0-as}\ ,
\end{equation}
with

\begin{eqnarray}K_{11}^{\phi}&=&K_{11}= -4(C_A+C_F)\label{Kphi11}\\
K_{10}^{\phi}&=&K_{10}=3C_F \label{Kphi10}\ .
\end{eqnarray}

b)\ \underbar{Two loop}:

\noindent 
Setting \cite{Soar:2009yh}:

\begin{eqnarray}c_{\phi g}^{(i)}(N)&\sim&\sum_{j=0}^{2i}c_{\phi g,j}^{(i)}\ln^j \bar{N} \label{iloop-cphig}\\
c_{\phi q}^{(i)}(N)&\sim&\frac{1}{N}\sum_{j=0}^{2i-1}c_{\phi q,j}^{(i)}\ln^{j} \bar{N} \label{iloop-cphiq}\ ,\end{eqnarray}
one gets for $i\geq 1$:

\begin{equation}K_{\phi}^{(i)}(N)\sim\sum_{j=0}^{2i} K_{i+1 j}^{\phi}\ln^j \bar{N}\label{Kphii-as}\ .
\end{equation}
At two loop one finds:
 \begin{eqnarray}K_{22}^{\phi}&=&\frac{1}{2}\tilde{\Delta}_{22}\beta_0-\Big[2 c_{\phi g,2}^{(1)}-\frac{1}{2}\tilde{c}_{\phi q,1}^{(1)}(A_1^g-A_1^q)\Big]\beta_0\label{Kphi22}\\
 K_{21}^{\phi}&=&\frac{1}{2}\tilde{\Delta}_{21}\beta_0-(A_2^g+A_2^q)\label{Kphi21}\\
 & &-\Big[2 c_{\phi g,1}^{(1)}-\frac{1}{2}\tilde{c}_{\phi q,0}^{(1)}(A_1^g-A_1^q)+\frac{1}{2}\tilde{c}_{\phi q,1}^{(1)}(B_1^g-B_1^q-\beta_0)\Big]\beta_0\nonumber\\
  K_{20}^{\phi}&=&\frac{1}{2}\tilde{\Delta}_{20}\beta_0+(B_2^g+B_2^q)-\beta_1\label{Kphi20}\\
 & &-\Big[2 c_{\phi g,0}^{(1)}+\frac{1}{2}\tilde{c}_{\phi q,0}^{(1)}(B_1^g-B_1^q-\beta_0)\Big]\beta_0\nonumber\ ,
 \end{eqnarray}
 where:
 
  \begin{eqnarray}\tilde{\Delta}_{ji}&= &\Delta_{ji}/C_F=-2\Delta_{ji}/\Delta_{10}\label{norm-Delta}\\
  \tilde{c}_{\phi q,i}^{(j)}&=&c_{\phi q,i}^{(j)}/C_F=-2c_{\phi q,i}^{(j)}/\Delta_{10}\label{norm-cphiq}\ .\end{eqnarray} 
  I note that the above results are obtained from those of section 3 simply by interchanging the ``renormalized'' off-diagonal parameters, i.e. $\tilde{D}_{ji}$'s with the  $\tilde{\Delta}_{ji}$'s, as well as the (``renormalized'') photon-exchange gluon coefficients with the (``renormalized'') $\phi$-exchange quark ones, and also  the diagonal photon-exchange quark coefficients with the $\phi$-exchange gluon ones, consistently with the remarks at the beginning of section 5.1.
  Using   known results on one loop coefficient functions and two loop anomalous dimensions  (see e.g.   \cite{Moch:2004pa,Vogt:2004mw,Soar:2009yh} and references therein), one obtains:

\begin{eqnarray}K_{22}^{\phi}&=& -2(C_A+C_F)\beta_0\label{Kphi22-value}\\
K_{21}^{\phi}&=&\Big(-\frac{16}{3}+8\zeta_2\Big)C_A^2 +\Big(-\frac{16}{3}+8\zeta_2\Big)C_A C_F -\frac{2}{3}C_A\beta_0-\frac{47}{3}C_F\beta_0-2\beta_0^2\label{Kphi21-value}\\
K_{20}^{\phi}&=&(-4+12\zeta_3)C_A^2 +(-10-12\zeta_3)C_A C_F+\Big(\frac{3}{2}-12\zeta_2+24\zeta_3\Big)C_F^2 \nonumber\\
& &+\Big(-\frac{10}{3}+4\zeta_2\Big) C_A\beta_0+\Big(17+6\zeta_2\Big)C_F\beta_0-\frac{29}{3}\beta_0^2-\beta_1 \label{Kphi20-value}\ .
\end{eqnarray}

c)\ \underbar{Three loop}:

\noindent 
At three loop, one gets:

 \begin{eqnarray}K_{3i}^{\phi}&=&\tilde{\Delta}_{3i}\beta_0+(lower\ loop\  quantities)\ (2\leq i\leq 4)\label{Kphi3i}\\
 K_{31}^{\phi}&=&\tilde{\Delta}_{31}\beta_0-(A_3^g+A_3^q)+(lower\ loop\  quantities) \label{Kphi31}\\
 K_{30}^{\phi}&=&\tilde{\Delta}_{30}\beta_0+(B_3^g+B_3^q)-\beta_2+(lower\ loop\  quantities) \label{Kphi30}\ .
 \end{eqnarray}
Using results on two loop coefficient functions  \cite{Soar:2009yh}  and three loop anomalous dimensions \cite{Moch:2004pa,Vogt:2004mw}, one finds:

\begin{eqnarray}K_{34}^{\phi}&=& 0\label{Kphi34-value}\\
K_{33}^{\phi}&=& -\frac{4}{3}(C_A+C_F)\beta_0^2\label{Kphi33-value}\\
K_{32}^{\phi}&=&\Big(\frac{44}{3}-24\zeta_2\Big)C_A^2\beta_0 +\Big(-\frac{136}{3}+72\zeta_2\Big)C_A C_F\beta_0 +(20-32\zeta_2) C_F^2\beta_0-\frac{2}{3}C_A\beta_0^2-\frac{47}{3}C_F\beta_0^2\nonumber\\
& &-2\beta_0^3-2( C_A+ C_F)\beta_1\label{Kphi32-value}\\
K_{31}^{\phi}&=&\Big(-\frac{436}{9}+\frac{80}{3}\zeta_2-96\zeta_3\Big)C_A^2\beta_0 +\Big(-\frac{1597}{9}-\frac{28}{3}\zeta_2+192\zeta_3\Big)C_A C_F\beta_0 +(-6+48\zeta_2-48\zeta_3) C_F^2\beta_0\nonumber\\
& &+\Big(\frac{104}{9}+36\zeta_2\Big)C_A\beta_0^2+\Big(-\frac{676}{9}-12\zeta_2\Big)C_F\beta_0^2+6 C_A\beta_1-9 C_F\beta_1-4\beta_0\beta_1-\frac{20}{3}\beta_0^3\label{Kphi31-value}\\
& &+\Big(\frac{224}{9}+\frac{64}{3}\zeta_2-\frac{176}{5}\zeta_2^2+176\zeta_3\Big)C_A^3+\Big(\frac{2039}{9}+\frac{64}{3}\zeta_2-\frac{176}{5}\zeta_2^2\Big)C_A^2 C_F+\Big(\frac{605}{3}-176\zeta_3\Big)C_A  C_F^2\nonumber\ .\end{eqnarray}

d)\ \underbar{Four loop}:

\noindent 
Finally, I quote the four loop result:

\begin{equation} K_{4i}^{\phi}=\frac{3}{2}\tilde{\Delta}_{4i}\beta_0+(lower\ loop\  quantities)\ (2\leq i\leq 6)\label{Kphi4i}\ .\end{equation}

\subsection{Threshold resummation and four loop predictions}

Here again, one finds the analogue of the ansatz (\ref{K-conjecture}), namely:

\begin{equation}
\label{Kphi-conjecture}
K_{\phi}(x,Q^2)\sim \frac{{\cal J}_{\phi}\left((1-x)Q^2\right)}{(1-x)_+} \end{equation}
is falsified in an interesting way. Indeed, setting:

\begin{equation}{\cal J}_{\phi}(Q^2)=j_{\phi 1}a_s+j_{\phi 2}a_s^2+j_{\phi 3}a_s^3+...\label{jphi}\ ,\end{equation}
one determines:

\begin{equation}j_{\phi 1}=-K_{11}^{\phi}\label{jphi1}\ , \end{equation}

\begin{equation}j_{\phi 2}=-K_{21}^{\phi}\  ,\end{equation}
whereas the  coefficient of the  single $\frac{\ln(1-x)}{1-x}$ term occurring at ${\cal O}(a_s^3)$ in $K_{\phi}(x,Q^2)$ is found to be given by:

\begin{equation}2 K_{32}^{\phi}=-[j_{\phi 1}\beta_1+2 (j_{\phi 2}+\delta_{\phi}) \beta_0]\label{phi-discrepancy}\ ,\end{equation}
with a ``discrepancy''  $\delta_{\phi}$ with respect to standard threshold resummation:

\begin{equation}\delta_{\phi}=4(-5+8\zeta_2)(C_A-C_F)^2\label{deltaphi-2}\ .\end{equation}
The interesting new feature of eq.(\ref{deltaphi-2}) compared to eq.(\ref{delta2}) is that the discrepancy is now {\em quadratic} in $C_{AF}\equiv C_A-C_F$. As we shall see, this feature implies that the predictions following from the ansatz eq.(\ref{Kphi-conjecture}) are correct  at ${\cal O}(a_s^3)$ not only for $C_A=C_F$, but also up to terms {\em linear} in $C_{AF}$. Indeed one obtains (as in the case of the $D_{ij}$'s) the correct \cite{Vogt:2004mw} two and three loop predictions:

\begin{equation}\Delta_{22}=-4C_F C_{AF}\label{Delta22predict}\end{equation}
and:

\begin{eqnarray}\Delta_{34}&=&-\frac{4}{3}C_FC_{AF}^2\label{Delta34predict}\\
\Delta_{33}&=&\frac{2}{3} C_F C_{AF}(5\beta_0-15C_F+12 C_{AF})\label{Delta33predict}\ Ê.\end{eqnarray}
Moreover one also  predicts:

\begin{eqnarray}\Delta_{32}\vert{predict}&=&C_F\Big[-3\beta_0^2+18 C_F\beta_0-27C_F^2\nonumber\\
& &-\frac{62}{3}\beta_0 C_{AF}-\Big(\frac{10}{3}-16\zeta_2\Big)C_F C_{AF}-\mathbf{\Big(\frac{76}{3}-32\zeta_2\Big)}C_{AF}^2\Big]\label{Delta32predict}\ Ê,\end{eqnarray}
while the exact result \cite{Vogt:2004mw} is:

\begin{eqnarray}\Delta_{32}&=&C_F\Big[-3\beta_0^2+18 C_F\beta_0-27C_F^2\nonumber\\
& &-\frac{62}{3}\beta_0 C_{AF}-\Big(\frac{10}{3}-16\zeta_2\Big)C_F C_{AF}-\mathbf{\frac{16}{3}}C_{AF}^2\Big]\label{Delta32}\ Ê,\end{eqnarray}
where the  mismatch between eq.(\ref{Delta32predict}) and eq.(\ref{Delta32}) has been highlighted in boldface. Indeed it is seen that the mismatch concerns only the term {\em quadratic} in $C_{AF}$. 

Finally, using known results on three loop coefficient functions \cite{Soar:2009yh}, the four loop predictions following from  leading logarithmic threshold resummation of $K_{\phi}(x,Q^2)$ are:

\begin{equation}
\Delta_{46}=0\label{Delta46predict}\ ,
\end{equation}

\begin{equation}
\Delta_{45}=\frac{2}{9} C_F C_{AF}^2(\beta_0-6 C_F+8C_{AF})\label{Delta45predict}\ ,
\end{equation}
and:

\begin{equation}\Delta_{44}=-C_F  C_{AF}\Big[\frac{8}{9}\beta_0^2-\frac{41}{6}C_F\beta_0+\frac{25}{2}C_F^2+\frac{208}{27}\beta_0 C_{AF}+\Big(\frac{224}{27}-8\zeta_2\Big)C_F C_{AF}+\Big(\frac{8}{27}+\frac{16}{9}\zeta_2\Big)C_{AF}^2 \Big]
\label{Delta44predict}\ ,\end{equation}
which have been first obtained in \cite{Soar:2009yh}, and   shown to be correct in \cite{Vogt:2010cv,Almasy:2010wn}. 

Moreover, assuming the validity of the ansatz (\ref{Kphi-conjecture}) also at the NLL level, one obtains the additional prediction:

\begin{eqnarray}
\Delta_{43}\vert{predict}&=&C_F\Big[\frac{4}{3}\beta_0^3-\frac{37}{3}C_F\beta_0^2+38 C_F^2\beta_0+\Big(-\frac{78037}{27}+\frac{15136}{9}\zeta_2+\frac{2048}{3}\zeta_3\Big)C_F^3\nonumber\\
& &+\frac{268}{27}\beta_0^2 C_{AF}+\Big(\frac{365}{27}-\frac{64}{3}\zeta_2\Big)\beta_0 C_F C_{AF}+\Big(-8677+\frac{15328}{3}\zeta_2+2048\zeta_3\Big)C_F^2 C_{AF}\nonumber\\
& &+C_{AF}^2 R_1(\beta_0,C_A,C_F)\Big]\label{Delta43predict}
\end{eqnarray}
where the coefficient $R_1$ of the  term quadratic in $C_{AF}$ is a linear homogeneous polynomial in $(C_A,C_F,\beta_0)$ which is not expected to be correctly predicted (at the difference of the coefficients of the terms {\em linear} in $C_{AF}$). Finally, the ansatz (\ref{Kphi-conjecture}) at the NNLL level yields a prediction for $\Delta_{42}$.  The latter however depends on the three loop non-logarithmic constant term $\Delta_{30}$, which could in principle be derived from the results in \cite{Vogt:2004mw}. Here I  only quote one important meaningful feature of the ensuing prediction: 

\begin{equation}
\Delta_{42}\vert{predict}=C_F P_3(\beta_0,C_A,C_F)+\frac{2}{3}C_F\frac{\beta_1}{\beta_0}\delta_{\phi}\label{Delta42predict}\ ,
\end{equation}
where $P_3$ is an  homogeneous cubic polynomial in $(C_A,C_F,\beta_0)$ (as in eq.(\ref{Delta43predict})), and the non-polynomial   contribution is {\em quadratic} in $C_{AF}$ (see eq.(\ref{deltaphi-2})).  Eq.(\ref{Delta42predict}) is thus {\em consistent} with the assumption that the ansatz (\ref{Kphi-conjecture}) is correct up to terms \emph{linear} in $C_{AF}$.

\section{Singlet fragmentation functions in $e^+e^-$ annihilation}

\subsection{Physical kernels}
The (transverse) singlet fragmentation function \cite{Cacciari:2001cw,Moch:2009my,Moch:2009hr,Blumlein:2000wh,Rijken:1996vr,Rijken:1996ns,Rijken:1996npa,Mitov:2006wy,Mitov:2006ic,Moch:2007tx,Stratmann:1996hn}  $F_T(N,Q^2)$ satisfies a short distance representation analogous  to eq.(\ref{OPE}):

\begin{equation} F_T(N,Q^2)= <e^2_q> \Big( C_q^T(N,Q^2/\mu^2,a_s) q_T(N,\mu^2)+C_g^T(N,Q^2/\mu^2,a_s) g_T(N,\mu^2)\Big)\label{short-distance}\ ,\end{equation}
where $C_a^T(N,Q^2/\mu^2,a_s)$  ($a=q,g$) are the singlet quark and gluon timelike coefficient functions, $q_T(N,\mu^2)=\sum_{i=1}^{n_f}(q_{iT}+\bar{q}_{iT})$ the singlet quark fragmentation distribution and $g_T(N,\mu^2)$ the gluon fragmentation    distribution.
The corresponding physical evolution kernels  are defined as in eq.(\ref{physkernels}):

\begin{equation}\ddot {F_T}(N,Q^2)=K^T(N,Q^2)\dot {F_T}(N,Q^2)+J^T(N,Q^2)F_T(N,Q^2)\label{physkernels-T}\ .
\end{equation}
In fact, all equations of section 2 take a similar form in the timelike case, except the renormalization group equations (\ref{dCq})-(\ref{ddCg}), where the indices $(qg)$ and $(gq)$ should be permuted, namely:

\begin{equation} \Big( \begin{array}{c} \!\! \dot{q_T}(N,\mu^2) \!\! \\ \! \dot{g_T}(N,\mu^2) \!\!
        \end{array} \Big) \:=\:
- \Big( \! \begin{array}{cc}\gamma_{qq}^T(N,a_s)\! & \gamma_{gq}^T(N,a_s)\! \\
  \gamma_{qg}^T(N,a_s)\! & \gamma_{gg}^T(N,a_s)\! \end{array}\! \Big) \Big( \begin{array}{c} \!\! q_T(N,\mu^2)\!\! \\ \! g_T(N,\mu^2) \!\!  \end{array} \Big)
\label{A-P-T}\ ,  \end{equation}
hence:

\begin{equation}\dot {C}_q^T(N,Q^2/\mu^2,a_s)=\beta(a_s)\frac{\partial C_q^T(N,Q^2/\mu^2,a_s)}{\partial a_s}-\gamma_{qq}^T(N,a_s)C_q^T(N,Q^2/\mu^2,a_s)-\gamma_{qg}^T(N,a_s)C_g^T(N,Q^2/\mu^2,a_s)\ ,\label{dCqT}\end{equation}

\begin{equation}\dot {C}_g^T(N,Q^2/\mu^2,a_s)=\beta(a_s)\frac{\partial C_g^T(N,Q^2/\mu^2,a_s)}{\partial a_s}-\gamma_{gq}^T(N,a_s)C_q^T(N,Q^2/\mu^2,a_s)-\gamma_{gg}^T(N,a_s)C_g^T(N,Q^2/\mu^2,a_s)\ ,\label{dCgT}
\end{equation}
and also:
\begin{equation}\ddot {C}_q^T(N,Q^2/\mu^2,a_s)=\beta(a_s)\frac{\partial \dot{C}_q^T(N,Q^2/\mu^2,a_s)}{\partial a_s}-\gamma_{qq}^T(N,a_s)\dot{C}_q^T(N,Q^2/\mu^2,a_s)-\gamma_{qg}^T(N,a_s)\dot{C}_g^T(N,Q^2/\mu^2,a_s)\ ,\label{ddCqT}\end{equation}

\begin{equation}\ddot {C}_g^T(N,Q^2/\mu^2,a_s)=\beta(a_s)\frac{\partial \dot{C}_g^T(N,Q^2/\mu^2,a_s)}{\partial a_s}-\gamma_{gq}^T(N,a_s)\dot{C}_q^T(N,Q^2/\mu^2,a_s)-\gamma_{gg}^T(N,a_s)\dot{C}_g^T(N,Q^2/\mu^2,a_s)\ .\label{ddCgT}
\end{equation}

\subsection{Large-$N$ Gribov-Lipatov relations}
The timelike coefficient functions are presently known up to  two loop \cite{Rijken:1996vr,Rijken:1996ns,Rijken:1996npa,Mitov:2006wy}, while
the timelike singlet anomalous dimensions are  known  up to two loop for the off-diagonal elements \cite{Curci:1980uw,Furmanski:1980cm,Floratos:1981hs}, and up to three loop for the diagonal ones \cite{Mitov:2006ic,Moch:2007tx}. Therefore the  physical kernels $K^T$ and $J^T$ can presently be computed only up to two loop \cite{Blumlein:2000wh}.

\noindent The results for the  $K^T(N,Q^2)$ kernel at large $N$ can be very simply summarized in the form of  large-$N$ Gribov-Lipatov like relations. Setting:

\begin{equation}K^T(N,Q^2)=\sum_{i=0}^{\infty}K^{(i)T}(N) a_s^{i+1}
\label{KT-expand}\ ,\end{equation}
one finds at large $N$:

a)\ \underbar{One loop}:

\begin{equation}K^{(0)T}(N)\sim K_{11}^T\ln \bar{N}+K_{10}^T\ ,\label{K0T-as}
\end{equation}
with:

\begin{equation}K_{1i}^T=K_{1i}\label{GL-1i}\ .
\end{equation}
This result is not surprising, since it is a straightforward consequence (see eqs.(\ref{K11}) and (\ref{K10})) for $a=b$ ($a,b=q,g$) of the well-known one loop Gribov-Lipatov relation \cite{Gribov:1972ri}:

\begin{equation}\gamma_{ab}^{(0)T}(N)=\gamma_{ab}^{(0)}(N)\ .\label{GL}
\end{equation}

b)\ \underbar{Two loop}:
More interestingly, although a Gribov-Lipatov like relation is known not to be valid at two loop at finite $N$, neither for the anomalous dimensions  \cite{Curci:1980uw,Stratmann:1996hn}, nor for the physical evolution kernels  \cite{Blumlein:2000wh}, one finds it does hold for the latter at large $N$. Indeed, setting for $N\rightarrow\infty$:

\begin{equation}K^{(1)T}(N)\sim\sum_{i=0}^2 K_{2i}^T\ln^i \bar{N}\ ,\label{K1T-as}
\end{equation}
one gets, using the results in  \cite{Rijken:1996vr,Rijken:1996ns,Rijken:1996npa,Mitov:2006wy}  and \cite{Curci:1980uw,Furmanski:1980cm,Floratos:1981hs}:

\begin{equation}K_{2i}^T= K_{2i}\  (i=1,2)\label{GL-2i}\ ,
\end{equation}
with:

 \begin{eqnarray}K_{22}^T&=&  \frac{1}{2}\tilde{\Delta}^T_{22}\beta_0-\Big[2 c_{q2}^{(1)T}-\frac{1}{2}\tilde{c}_{g1}^{(1)T}(A_1^{gT}-A_1^{qT})\Big]\beta_0\label{K22T}\\
 K_{21}^T&=& \frac{1}{2}\tilde{\Delta}^T_{21}\beta_0-(A_2^{gT}+A_2^{qT})\label{K21T}\\
 & &-\Big[2 c_{q1}^{(1)T}-\frac{1}{2}\tilde{c}_{g0}^{(1)T}(A_1^{gT}-A_1^{qT})+\frac{1}{2}\tilde{c}_{g1}^{(1)T}(B_1^{gT}-B_1^{qT}-\beta_0)\Big]\beta_0\nonumber\\
  K_{20}^T&=&\frac{1}{2}\tilde{\Delta}^T_{20}\beta_0+(B_2^{gT}+B_2^{qT})-\beta_1\label{K20T}\\
 & &-\Big[2 c_{q0}^{(1)T}+\frac{1}{2}\tilde{c}_{g0}^{(1)T}(B_1^{gT}-B_1^{qT}-\beta_0)\Big]\beta_0\nonumber\ .
 \end{eqnarray}
The $\tilde{\Delta}^T_{ij}$'s are defined by:

 \begin{equation}\tilde{\Delta}^T_{ij}= \Delta^T_{ij}/C_F=-2 \Delta^T_{ij}/\Delta^T_{10}\label{Delta-norm}\ ,\end{equation}
 with: 

\begin{equation}\gamma_{gq}^{(i)T}(N)\sim\frac{1}{N}\sum_{j=0}^{2i} \Delta^T_{i+1 j}\ln^j\bar{N}\label{gammaTgq-as}\ ,\end{equation}
($i\geq 0$), and

  \begin{equation}\tilde{c}_{gj}^{(i)T}=c_{gj}^{(i)T}/C_F=-2c_{gj}^{(i)T}/\Delta^T_{10}\label{cgT-norm}\ ,\end{equation}
 with:  
 
\begin{equation} c_g^{(i)T}(N)\sim\frac{1}{N}\sum_{j=0}^{2i-1} c_{gj}^{(i)T}\ln^j \bar{N} \label{jloop-cgT}\ ,\end{equation}
($i\geq 1$).
Moreover:

\begin{equation} c_q^{(i)T}(N)\sim\sum_{j=0}^{2i} c_{qj}^{(i)T}\ln^j \bar{N} \label{jloop-cqT}\ .\end{equation}
Using eqs.(\ref{K22T}), (\ref{K22}) and (\ref{K21T}),  (\ref{K21}), eq.(\ref{GL-2i}) is seen to be equivalent to:

\begin{eqnarray}\tilde{\Delta}^T_{22}- \tilde{D}_{22}&=&0\label{GLbis-22}\\
 \tilde{\Delta}^T_{21}- \tilde{D}_{21}&=&-(\delta_{TS}\tilde{c}_{g0}^{(1)})(A_1^{g}-A_1^{q})\label{GLbis-21}\ ,
\end{eqnarray}
where $\delta_{TS}f\equiv f^T-f$, and  the  relations \cite{Moch:2007tx,Dokshitzer:2005bf,Basso:2006nk}: $\delta_{TS}A_j^{a}=\delta_{TS}B_j^{a}=0$ and \cite{Moch:2009my} $\delta_{TS}c_{q2}^{(1)}=\delta_{TS}c_{q1}^{(1)}=0$  have been used, together with\footnote{The normalization convention \cite{Vogt} adopted here (see eqs.(\ref{short-distance}), (\ref{A-P-T}) and (\ref{GL})) requires the timelike gluon coefficient function $C^T_g $ to be  $1/2$ the standard one in \cite{Rijken:1996vr,Rijken:1996ns,Rijken:1996npa,Mitov:2006wy} (and the timelike gluon distribution $g_T$ to be $2 n_f\times$ the standard  one).}:

\begin{equation}\delta_{TS}\tilde{c}_{g1}^{(1)}=0\label{delta-cg1}\ .\end{equation}

\noindent One also finds that a  relation similar to eq.(\ref{GL-2i}) does {\em not} hold for the $K_{20}^T$ constant term. Instead one gets:

\begin{equation}K_{20}^T-K_{20}=-12 \zeta_2 (C_A+C_F)\beta_0\label{GL-20}\ .
\end{equation}
Eq.(\ref{GL-20}) means that in momentum space  only the $\frac{1}{(1-x)_+}$ part of $K^{(1)T}(x)$ satisfies a Gribov-Lipatov relation, not the $\delta(1-x)$ part, namely we have for $x\rightarrow 1$:

\begin{eqnarray}
K^{(1)}(x)&\sim&\frac{2 K_{22}\ln (1-x)-K_{21}}{(1-x)_+}+(K_{20}-\zeta_2 K_{22})\delta(1-x) \label{K1-x-large}\\
K^{(1)T}(x)&\sim&\frac{2 K_{22}\ln (1-x)-K_{21}}{(1-x)_+}+(K_{20}^T-\zeta_2 K_{22})\delta(1-x) \label{K1T-x-large}\  .\end{eqnarray}
Similar large-$x$ relations\footnote{Similar relations hold \cite{Moch:2007tx,Dokshitzer:2005bf,Basso:2006nk} for the $1/(1-x)_+$ parts of {\em diagonal} splitting functions;  in this case, the $\delta(1-x)$ terms are also equal.} have been observed \cite{Grunberg:2009vs,Grunberg:2010sw} up to $i=2$  for the non singlet physical evolution kernels $K^{(i)T}_{ns}$ and $K^{(i)}_{ns}$.

\subsection{Three loop timelike predictions}
Given the  non-trivial character of eqs.(\ref{GL-2i}) (see in particular eq.(\ref{GLbis-21})), it is natural to assume they are not accidental, and that similar large-$N$ Gribov-Lipatov like relations also hold beyond two loop, as in the non-singlet case  \cite{Grunberg:2009vs,Grunberg:2010sw}. At three loop one gets for $N\rightarrow\infty$:

 \begin{equation}K^{(2)T}(N)\sim \sum_{i=0}^4 K^T_{3i}\ln^i \bar{N}\label{KT2-as}
\end{equation}
with:

 \begin{eqnarray}K^T_{3i}&=&\tilde{\Delta}^T_{3i}\beta_0+(lower\ loop\  quantities)\ (2\leq i\leq 4)\label{KT3i}\\
 K^T_{31}&=&\tilde{\Delta}^T_{31}\beta_0-(A_3^{gT}+A_3^{qT})+(lower\ loop\  quantities) \label{KT31}\\
 K^T_{30}&=&\tilde{\Delta}^T_{30}\beta_0+(B_3^{gT}+B_3^{qT})-\beta_2+(lower\ loop\  quantities) \label{KT30} \ ,
  \end{eqnarray}
 where the ``lower loop quantities'' do not depend on the off-diagonal $\gamma^T_{qg}(N)$ anomalous dimension, i.e. on the $D^T_{ij}$'s.
Assuming the large-$N$ Gribov-Lipatov  relations:

\begin{equation}K_{3i}^T= K_{3i}\  (1\leq i\leq 4)\label{GL-3i}\ ,
\end{equation}
eqs.(\ref{KT3i}), (\ref{K3i}) and (\ref{KT31}), (\ref{K31}) then show that the $\tilde{\Delta}^T_{3i}$'s ($1\leq i\leq 4$) can be predicted in term of the $\tilde{D}_{3i}$'s (and three loop cusp anomalous dimensions) and lower loop quantities. Indeed eq.(\ref{GL-3i})  yields:

\begin{eqnarray}\tilde{\Delta}^T_{34}- \tilde{D}_{34}&=&0\label{GLbis-34}\\
 \tilde{\Delta}^T_{33}- \tilde{D}_{33}&=&(A_1^{g}-A_1^{q})\Big[-(\delta_{TS}\tilde{c}_{g2}^{(2)}) +(\delta_{TS}\tilde{c}_{g0}^{(1)})c_{q2}^{(1)}\Big]\label{GLbis-33}\\
 \tilde{\Delta}^T_{32}- \tilde{D}_{32}&=&-\frac{1}{4}(A_1^{g}-A_1^{q})^2 (\delta_{TS}\tilde{c}_{g0}^{(1)}) (\tilde{c}_{g0}^{(1)T}+\tilde{c}_{g0}^{(1)})\nonumber\\
 & &+\frac{1}{4}(A_1^{g}-A_1^{q}) (\delta_{TS}\tilde{c}_{g0}^{(1)})(\tilde{\Delta}^T_{21}+\tilde{D}_{21})+4(\delta_{TS}c_{q2}^{(2)})\nonumber\\
& & -(A_1^{g}-A_1^{q})(\delta_{TS}\tilde{c}_{g1}^{(2)})+(B_1^{g}-B_1^{q}-2\beta_0)(\delta_{TS}\tilde{c}_{g2}^{(2)})\nonumber\\
& &-\frac{1}{2}(\tilde{\Delta}^T_{20}- \tilde{D}_{20})\Big[\tilde{D}_{22}+(A_1^{g}-A_1^{q})\tilde{c}_{g1}^{(1)}\Big]\nonumber\\
& &+\Big[(A_1^{g}-A_1^{q})\tilde{c}_{g1}^{(1)}-4 c_{q2}^{(1)}\Big](\delta_{TS}c_{q0}^{(1)})\nonumber\\
& &-\frac{1}{2}(A_1^{g}-A_1^{q})(\delta_{TS}\tilde{c}_{g0}^{(1)})(B_1^{g}-B_1^{q}-\beta_0)\tilde{c}_{g1}^{(1)}\nonumber\\
& &-\frac{1}{2}(A_1^{g}-A_1^{q})(\tilde{\Delta}^T_{21}\tilde{c}_{g0}^{(1)T}-\tilde{D}_{21} \tilde{c}_{g0}^{(1)})\nonumber\\
& &+(\delta_{TS}\tilde{c}_{g0}^{(1)})\Big[\frac{1}{2}\tilde{D}_{22}(B_1^{g}-B_1^{q}-\beta_0)+(A_1^{g}-A_1^{q})(B_1^{g}-B_1^{q})\tilde{c}_{g1}^{(1)}\nonumber\\
& &+(A_1^{g}-A_1^{q})c_{q1}^{(1)}- (B_1^{g}-B_1^{q}-2\beta_0)c_{q2}^{(1)}-\beta_0(A_1^{g}-A_1^{q})\tilde{c}_{g1}^{(1)}\Big]
 \label{GLbis-32}\end{eqnarray}
(with a similar, even more lengthy equation for $\tilde{\Delta}^T_{31}- \tilde{D}_{31}$, not written down for brevity),
where I used \cite{Moch:2009my} $\delta_{TS}c_{q3}^{(2)}=\delta_{TS}c_{q4}^{(2)}=0$, and also  eqs.(\ref{GLbis-22}) and (\ref{GLbis-21}), together with:

\begin{equation}\delta_{TS}\tilde{c}_{g3}^{(2)}=0\label{delta-cg2}\ .\end{equation}
Thus the large-$N$ Gribov-Lipatov relations eq.(\ref{GL-3i}) determines the $\Delta^T_{3i}$'s  in term of the $D_{3i}$'s and lower loop quantities.
One   finds:

\begin{eqnarray}\Delta^T_{34}&=&-\frac{4}{3}C_F C_{AF}^2\label{DeltaT-34}\\
\Delta^T_{33}&=&\frac{2}{3}C_F C_{AF}(12C_A-9C_F-\beta_0)\label{DeltaT-33}\\
\Delta^T_{32}&= & \frac{2}{3}C_F C_{AF}[(8-24\zeta_2)C_A+51 C_F+13 \beta_0)]\label{DeltaT-32}\\
\Delta^T_{31}&= &\Big(\frac{214}{9}-10\zeta_2\Big)C_F^2\beta_0+\Big(-\frac{196}{9}+14\zeta_2\Big)C_A C_F\beta_0+\Big(\frac{278}{9}+46\zeta_2-\frac{560}{3}\zeta_3\Big)C_A C_F^2\nonumber\\
& &+\Big(-\frac{206}{9}-32\zeta_2+\frac{208}{3}\zeta_3\Big)C_A^2 C_F+\Big(-14-26\zeta_2+\frac{352}{3}\zeta_3\Big) C_F^3\label{DeltaT-31}\ .\end{eqnarray}
Inverting to momentum space, one gets for $x\rightarrow 1$:

\begin{equation}P_{gq}^{(2)T}(x)\sim\sum_{i=0}^{4} \bar{\Delta}^T_{3 i}\ln^i(1-x)\label{gammaTgq(x)-as}\ ,\end{equation}
with:

\begin{eqnarray}\bar{\Delta}^T_{34}&=&\frac{4}{3}C_F C_{AF}^2\label{DeltabarT-34}\\
\bar{\Delta}^T_{33}&=&\frac{2}{3}C_F C_{AF}(12C_A-9C_F-\beta_0)\label{DeltabarT-33}\\
\bar{\Delta}^T_{32}&= & -\frac{2}{3}C_F C_{AF}[(8-12\zeta_2)C_A+(51-12\zeta_2) C_F+13 \beta_0)]\label{DeltabarT-32}\\
\bar{\Delta}^T_{31}&= &\Big(\frac{214}{9}-12\zeta_2\Big)C_F^2\beta_0+\Big(-\frac{196}{9}+16\zeta_2\Big)C_A C_F\beta_0+\Big(\frac{278}{9}+88\zeta_2-208\zeta_3\Big)C_A C_F^2\nonumber\\
& &+\Big(-\frac{206}{9}-56\zeta_2+80\zeta_3\Big)C_A^2 C_F+\Big(-14-44\zeta_2+128\zeta_3\Big) C_F^3\label{DeltabarT-31}\ .\end{eqnarray}
Of course, the double logarithmic terms $\Delta^T_{3i}\ (i=3,4)$ could also have been obtained from an alternative ansatz similar to the one used in section 4.2. The present Gribov-Lipatov ansatz however  yields in addition the single logarithmic terms $\Delta^T_{3i}\ (i=1,2)$. I note the predicted $\Delta^T_{3i}$'s vanish when $C_A=C_F$ for $i=2,3,4$, similarly to the $D_{3i}$'s.

\noindent The Gribov-Lipatov relations eq.(\ref{GL-3i}) are not expected to be valid for the constant terms ($i=0$). Nevertheless, it is interesting to speculate about a simple structure of the difference $K_{30}^T-K_{30}$, similar to the one observed in the two loop case,  eq.(\ref{GL-20}). Indeed one finds:

\begin{eqnarray}K_{30}^T-K_{30}&=&
-12 \zeta_2 (C_A+C_F)\beta_1+[\tilde{\Delta}^T_{30}- \tilde{D}_{30}]\beta_0\label{GL-30}\\
& &+\Big[(-6+48 \zeta_2+144 \zeta_2^2-24\zeta_3)C_A^2+(8+44 \zeta_2-240 \zeta_2^2-24\zeta_3)C_A C_F\nonumber\\
& &+(18-186 \zeta_2+192 \zeta_2^2+48\zeta_3)C_F^2\big]\beta_0+\Big[(4-12 \zeta_2)C_A+(6-134 \zeta_2)C_F\Big]\beta_0^2\nonumber\ ,
\end{eqnarray}
where all explicitly calculated contributions arise from less then three loop, and $\tilde{D}_{30}$ could be in principle extracted from the results in \cite{Vogt:2004mw}. The first term on the right-hand side of eq.(\ref{GL-30}) reveals that indeed the difference $K_{30}^T-K_{30}$ cannot vanish, since this term cannot be canceled by the other contributions, which are all proportional  to $\beta_0$. However, as suggested by the right-hand side of eq.(\ref{GL-20}), a number of cancellations may take place. In particular, one might expect that all  terms with  no factor of  $\zeta_2$ cancel on the right-hand side of eq.(\ref{GL-30}):

\begin{equation}K_{30}^T-K_{30}={\cal O}(\zeta_2)\label{assumption1}\ ,\end{equation}
a possibility which would be interesting to check against the forthcoming exact result \cite{Vogt}.

It is clear that similar methods may be used to predict the large-$x$  logarithmic coefficients ${\bar D}_{3i}^T$ ($1\leq i\leq 4$) in the  three loop off-diagonal timelike splitting function $P_{qg}^{(2)T}(x)$:

\begin{equation}P_{qg}^{(2)T}(x)\sim\sum_{i=0}^{4} \bar{D}^T_{3 i}\ln^i(1-x)\label{gammaTqg(x)-as}\ ,\end{equation}
assuming the analogue of the large-$N$ Gribov-Lipatov relations eq.(\ref{GL-3i}) for the $\phi$-related physical kernels:

\begin{equation}K_{3i}^{\phi T}= K_{3i}^{\phi}\  (1\leq i\leq 4)\label{phiGL-3i}\ .
\end{equation}
Eq.(\ref{phiGL-3i}) yields relations between the $ D_{3i}^T$'s and the $ \Delta_{3i}$'s, which however depend on the 
 one and two loop $\phi$-decay timelike  quark and gluon coefficient functions. The latter could in principle be determined \cite{Moch:2007tx} by analytic continuation of  corresponding spacelike quantities. 

\section{Conclusion}
The large-$x$/ large-$N$ behavior of  the  physical evolution kernels associated to the second order evolution equations for the singlet $F_2$ (photon-exchange DIS) and $F_{\phi}$ ($\phi$-exchange DIS) structure functions has been investigated. It was shown that for each process at {\em large} $N$ there is actually only {\em one} independent ``scalar'' physical kernel $K(x,Q^2)$  (resp. $K_{\phi}(x,Q^2)$) for  singlet evolution, in addition to the non-singlet one $K_{ns}(x,Q^2)$ (resp.  $K_{\phi,ns}(x,Q^2)$). 

\noindent The singlet kernel $K(x,Q^2)$ was found to satisfy  up to three loop a leading logarithmic standard form of threshold resummation, analogous to the one valid for the non-singlet kernel, and which holds presumably also beyond three loop. This assumption allows to predict the double logarithmic contributions to the four loop off-diagonal splitting function $P^{(3)}_{qg}(x)$, thus recovering some of the results of \cite{Soar:2009yh}. It was shown that this agreement is not accidental, and that the assumption of leading logarithmic threshold resummation of the singlet ``scalar'' kernel $K(x,Q^2)$  is closely related to the assumption (now confirmed \cite{Vogt:2010cv,Almasy:2010wn}) of single logarithmic behavior of the ``matrix'' physical kernel studied in  \cite{Soar:2009yh}. The  present more intrinsic ``scalar'' approach  allows however to go a little further, and provides some information on the two leading large $x$ {\em single logarithmic} contributions ($\ln^i(1-x)$, $i=2,3$) to $P^{(3)}_{qg}(x)$. It is found  that there is an obstruction to threshold resummation (as opposed to the non-singlet case) at the next-to-leading logarithmic level, which first manifests itself in the three loop kernel (a similar fact has been observed \cite{Moch:2009mu,Grunberg:2009am} in the case of the non-singlet longitudinal $F_L$ structure function \cite{Akhoury:1998gs,Akhoury:2003fw}). This obstruction turns out to be removed in the ``supersymmetric'' case $C_A=C_F$, as well as at large $\beta_0$, which suggests a full threshold resummation might still be possible in these cases, leading to the above mentioned additional predictions. On the other hand, no prediction is available for the ${\cal O}(\ln(1-x))$ term in $P^{(3)}_{qg}(x)$, which should be viewed as an {\em input} for the threshold resummation.
Of course, more work is needed to justify the assumed threshold resummation for $C_A=C_F$ (for which a non-trivial consistency check has been provided, see the comment after eq.(\ref{D42predict})), and to better understand the general $C_A\neq C_F$ situation.

\noindent  The analogous study of $\phi$-exchange DIS, where the scalar $\phi$ is directly coupled to gluons, gives information on the other off-diagonal splitting function $P_{gq}(x)$. The interesting new feature in this case is that the discrepancy at three loop with threshold resummation is found to be {\em quadratic} in  $C_A-C_F$, allowing to predict more terms in the  single logarithmic contributions to $P^{(3)}_{gq}(x)$.

\noindent A similar approach has been applied to the study of the large-$x$ behavior of $e^+ e^-$ fragmentation functions. A large-$x$ Gribov-Lipatov relation  (similar to the one holding \cite{Grunberg:2009vs,Grunberg:2010sw} for the non-singlet kernels)  has been observed in the two loop physical kernels. Assuming a similar relation is valid at three loop, {\em all} large-$x$ logarithmic contributions to the three loop timelike off-diagonal splitting function $P^{(2)T}_{gq}(x)$ are predicted and related to corresponding terms in $P^{(2)}_{qg}(x)$ , which could be checked against exact calculations \cite{Vogt} in the near future.

\vspace{0.5cm}

\noindent\textbf{Acknowledgements} I thank A. Vogt for a very helpful correspondence on the normalization conventions for the  timelike gluon coefficient function, and A. Mitov for raising a thought-stimulating question.

\vspace{0.3cm}

\appendix

\section{Connection with the matrix physical evolution kernel  for the $(F_2, F_{\phi})$ system}
An standard alternative  method \cite{Furmanski:1981cw}  to construct physical evolution kernels for singlet evolution consists in the simultaneous introduction of  another process in addition to the photon exchange one,  in particular the exchange of a scalar $\phi$ directly coupled to gluons. This system has been considered in \cite{Soar:2009yh}.
The matrix physical evolution kernel for the coupled $(F_2 ,F_{\phi})$ system  is defined by:

\begin{equation} \Big( \begin{array}{c} \!\! \dot{F}(N,Q^2) \!\! \\ \! \dot{F}_{\phi}(N,Q^2) \!\!
        \end{array} \Big) \:=\:
 \Big( \! \begin{array}{cc} K_{22}(N,Q^2)\! & K_{2\phi}(N,Q^2)\! \\
  K_{\phi 2}(N,Q^2)\! & K_{\phi\phi}(N,Q^2)\! \end{array}\! \Big) \Big( \begin{array}{c} \!\! F (N,Q^2)\!\! \\ \! F_{\phi}(N,Q^2) \!\!  \end{array} \Big)
\label{g-f-kernel}\ ,   \end{equation}
with $F$ as in eq.(\ref{F}).
Taking the $Q^2$ derivative of both sides of eq.(\ref{g-f-kernel}),  considering the first line (relevant for the photon exchange case), and eliminating $F_{\phi}$ and $\dot{F}_{\phi}$, one recovers the ``scalar'' evolution equation eq.(\ref{physkernels}) with the identifications:

\begin{eqnarray}K(N,Q^2)&=&K_{22}(N,Q^2)+K_{\phi\phi}(N,Q^2) +\frac{\dot{K}_{2\phi}(N,Q^2)}{K_{2\phi}(N,Q^2)}\label{physkernel-K-bis}\\
J(N,Q^2)&=&K_{2\phi}(N,Q^2) K_{\phi 2}(N,Q^2)-K_{22}(N,Q^2) K_{\phi\phi}(N,Q^2)\nonumber\\
& &+\dot{K}_{22}(N,Q^2) -K_{22}(N,Q^2)\frac{\dot{K}_{2\phi}(N,Q^2)}{K_{2\phi}(N,Q^2)}\label{physkernel-J-bis}\ .
\end{eqnarray}
Taking next the derivative of the second line of eq.(\ref{g-f-kernel}), and eliminating $F$ and $\dot{F}$, one   recovers similarly eq.(\ref{phi-physkernels}), with $K_{\phi}(N,Q^2)$ and  $J_{\phi}(N,Q^2)$ obtained by relations analogous to eqs.(\ref{physkernel-K-bis}) and (\ref{physkernel-J-bis}) with the indices $2$ and $\phi$ on the right-hand sides  interchanged.

\subsection{Relation between the $J$ (resp. $J_{\phi}$) and $K$ (resp. $K_{\phi}$) singlet physical kernels}
\noindent Consider now the large $N$ limit. I first observe that in this limit, $J(N,Q^2)$ can be expressed in term of $K(N,Q^2)$ and $K_{22}(N,Q^2)$. Indeed, it is known \cite{Soar:2009yh} that at large $N$, $K_{22}(N,Q^2)$ and $K_{\phi\phi}(N,Q^2)$ are ${\cal O}(N^0)$ (up to logarithms of $N$), while $K_{2\phi}(N,Q^2)$ and $K_{\phi2}(N,Q^2)$ are ${\cal O}(1/N)$. Thus for $N\rightarrow\infty$ one can neglect the first term on the right-hand side of eq.(\ref{physkernel-J-bis}), and get:

\begin{equation}J(N,Q^2)\sim {\dot K}_{22}(N,Q^2)-K_{22}(N,Q^2)\Big[K_{\phi\phi}(N,Q^2) +\frac{\dot{K}_{2\phi}(N,Q^2)}{K_{2\phi}(N,Q^2)}\Big]\label{J-largeN1}\ .
\end{equation}
Using eq.(\ref{physkernel-K-bis}), one then finds:

\begin{equation}J(N,Q^2)\sim {\dot K}_{22}(N,Q^2)-K_{22}(N,Q^2)\Big[K(N,Q^2)-K_{22}(N,Q^2)\Big]\label{J-largeN2}\ .
\end{equation}
Moreover, it follows from \ \cite{Soar:2009yh} that at large $N$, the ${\cal O}(N^0)$ part (modulo logarithms of $N$) of $K_{22}(N,Q^2)$ coincides with the non singlet kernel $K_{ns}(N,Q^2)$, which is known \cite{Gardi:2007ma,Friot:2007fd} to satisfy a standard form of threshold resummation. Indeed, in momentum space one can check that   eqs.(4.13)-(4.16) of  \cite{Soar:2009yh} can be cast in the form of eqs.(\ref{K-conjecture}) and (\ref{j-expand}) (truncated to ${\cal O}(a_s^3)$),  namely we have at large $x$:

\begin{equation}
\label{K22-conjecture}
K_{22}(x,Q^2)\sim \frac{{\cal J}_{ns}\left((1-x)Q^2\right)}{(1-x)_+}\sim K_{ns}(x,Q^2)\ , \end{equation}
where ${\cal J}_{ns}\left((1-x)Q^2\right)$ satisfies the analogue  of   eq.(\ref{j-expand}) (with $j_{ns,1}=4C_F$).
Thus we obtain:

\begin{equation}J(N,Q^2)\sim {\dot K}_{ns}(N,Q^2)-K_{ns}(N,Q^2)\Big[K(N,Q^2)-K_{ns}(N,Q^2)\Big]\label{J-largeN3}\ ,
\end{equation}
which, for $N\rightarrow\infty$, expresses\footnote{For $K(N,Q^2)=K_{ns}(N,Q^2)$, eq.(\ref{J-largeN3}) becomes $J(N,Q^2)\sim{\dot K}_{ns}(N,Q^2)$, which also follows by taking the $Q^2$ derivative of eq.(\ref{ns-physkernel}), and in this case is  valid at finite $N$ too.}  $J(N,Q^2)$ in term of  $K(N,Q^2)$ and $K_{ns}(N,Q^2)$ up to ${\cal O}(1/N^2)$ corrections. It thus suffices to study the large-$N$ properties of $K(N,Q^2)$.

Similarly, $J_{\phi}(N,Q^2)$ can be expressed at large $N$ in term of $K_{\phi}(N,Q^2)$ and $K_{\phi \phi}(N,Q^2)$:

\begin{equation}J_{\phi}(N,Q^2)\sim {\dot K}_{\phi \phi}(N,Q^2)-K_{\phi \phi}(N,Q^2)\Big[K_{\phi}(N,Q^2)-K_{\phi \phi}(N,Q^2)\Big]\label{Jphi-largeN2}\ .
\end{equation}
The results of \cite{Soar:2009yh} also imply that $K_{\phi \phi}(N,Q^2)$  satisfies  at large $N$  a standard non-singlet form of threshold resummation\footnote{This fact is closely related to the known \cite{Moch:2005ba,Soar:2009yh} soft gluon exponentiation of $C_{\phi,g}$.},  and can be identified to a non-singlet ``gluonic'' physical kernel. Indeed, one can check that  eqs.  (4.18)-(4.21)  of  \cite{Soar:2009yh} can also be cast in the form of eqs.(\ref{K-conjecture}) and (\ref{j-expand}) (truncated to ${\cal O}(a_s^3)$),  namely we have:
 
\begin{equation}
\label{Kphiphi-conjecture}
K_{\phi \phi}(x,Q^2)\sim \frac{{\cal J}_{\phi,ns}\left((1-x)Q^2\right)}{(1-x)_+}\sim K_{\phi,ns}(x,Q^2)\ , \end{equation}
where ${\cal J}_{\phi,ns}\left((1-x)Q^2\right)$ satisfies the analogue  of   eq.(\ref{j-expand}) (with $j_1^{\phi,ns}=4C_A$). 

\noindent To summarize, at large $N$ there are only two independent singlet ``scalar'' physical kernels ($K(N,Q^2)$ and $K_{\phi}(N,Q^2)$), and two independent non-singlet kernels ($K_{ns}(N,Q^2)$ and $K_{\phi,ns}(N,Q^2)$), which match the four elements of the matrix kernel (\ref{g-f-kernel}).

\subsection{Single logarithmic enhancement of $K_{2\phi}$ (resp. $K_{\phi 2}$) and leading logarithmic threshold resummation of $K$ (resp. $K_{\phi}$)}
Setting:

\begin{equation}K_{2\phi}(N,Q^2)=\sum_{i=0}^{\infty}K_{2\phi}^{(i)}(N) a_s^{i+1}
\label{K2phi-expand}\ ,\end{equation}
let us first assume  the  large-$N$ behavior of the off-diagonal kernel coefficients $K_{2\phi}^{(i)}(N)$ to be doubly logarithmic (as a priori expected from the behavior of the off-diagonal splitting functions):

\begin{equation}K_{2\phi}^{(i)}(N)\sim\frac{1}{N}\sum_{j=0}^{2i} D^{\phi}_{i+1 j}\ln^j\bar{N}\label{K2phi-as}\ .\end{equation}
On the other hand, for  the diagonal kernels coefficients  $K_{aa}^{(i)}(N)$  ($a=2,\phi$),  assuming a single logarithmic enhancement, as follows from  their large-$N$ identification with  non-singlet physical kernels, and their ensuing  threshold resummation properties (Appendix A.1), one has:

\begin{equation}K_{aa}(N,Q^2)=\sum_{i=0}^{\infty}K_{aa}^{(i)}(N) a_s^{i+1}
\label{Kaa-expand}\ ,\end{equation}
 with:

 \begin{eqnarray}K_{22}^{(i)}(N)&\sim&\sum_{j=0}^{i+1} K^{ns}_{i+1 j}\ln^j \bar{N}\label{K22-as}\\
 K_{\phi\phi}^{(i)}(N)&\sim&\sum_{j=0}^{i+1} K^{\phi, ns}_{i+1 j}\ln^j \bar{N}\label{Kphiphi-as}\ .
\end{eqnarray}
Using eqs.(\ref{K2phi-expand})-(\ref{Kphiphi-as}) into eq.(\ref{physkernel-K-bis}), one deduces the large-$N$ behavior (eqs.(\ref{K0-as}) and (\ref{Ki-as})) of the $K^{(i)}(N)$'s (eq.(\ref{K-expand})), namely:

i) At one loop:

 \begin{eqnarray}K_{11}&=& K^{ns}_{11}+K^{\phi, ns}_{11}\label{K11-bis}\\
 K_{10}&=&K^{ns}_{10}+K^{\phi, ns}_{10}-\beta_0 \label{K10-bis}\ .
\end{eqnarray}

ii) At two loop:

 \begin{eqnarray}K_{22}&=& K^{ns}_{22}+K^{\phi, ns}_{22}-\frac{1}{2}\tilde{D}^{\phi}_{22}\beta_0\label{K22-bis}\\
 K_{21}&=&K^{ns}_{21}+K^{\phi,ns}_{21}-\frac{1}{2}\tilde{D}^{\phi}_{21}\beta_0\label{K21-bis}\\
 K_{20}&=&K^{ns}_{20}+K^{\phi, ns}_{20}-\frac{1}{2}\tilde{D}^{\phi}_{20}\beta_0-\beta_1\label{K20-bis}\ ,
 \end{eqnarray}
where:

 \begin{equation}\tilde{D}^{\phi}_{i j}=D^{\phi}_{i j}/n_f=2 D^{\phi}_{i j}/D^{\phi}_{10}\label{D-phi-norm}\ .\end{equation}

iii) At three loop:

 \begin{eqnarray}K_{34}&=&\Big[-\tilde{D}^{\phi}_{34}+\Big(\frac{1}{2}\tilde{D}^{\phi}_{22}\Big)^2\Big]\beta_0\label{K34-bis}\\
 K_{33}&=&K^{ns}_{33}+K^{\phi, ns}_{33}+\Big[-\tilde{D}^{\phi}_{33}+\frac{1}{2}\tilde{D}^{\phi}_{22}\tilde{D}^{\phi}_{21}\Big]\beta_0\label{K33-bis}\\
 K_{32}&=&K^{ns}_{32}+K^{\phi, ns}_{32}+\Big[-\tilde{D}^{\phi}_{32}+\Big(\frac{1}{2}\tilde{D}^{\phi}_{21}\Big)^2+\frac{1}{2}\tilde{D}^{\phi}_{22}\tilde{D}^{\phi}_{20}\Big]\beta_0-\frac{1}{2}\tilde{D}^{\phi}_{22}\beta_1\label{K32-bis}\\
 K_{31}&=&K^{ns}_{31}+K^{\phi, ns}_{31}+\Big[-\tilde{D}^{\phi}_{31}+\frac{1}{2}\tilde{D}^{\phi}_{21}\tilde{D}^{\phi}_{20}\Big]\beta_0-\frac{1}{2}\tilde{D}^{\phi}_{21}\beta_1\label{K31-bis}\\
  K_{30}&=&K^{ns}_{30}+K^{\phi, ns}_{30}+\Big[-\tilde{D}^{\phi}_{30}+\Big(\frac{1}{2}\tilde{D}^{\phi}_{20}\Big)^2\Big]\beta_0-\frac{1}{2}\tilde{D}^{\phi}_{20}\beta_1-\beta_2\label{K30-bis}
 \ . \end{eqnarray}

iv) At four loop:

\begin{eqnarray}K_{46}&=&\Big[-\frac{3}{2}\tilde{D}^{\phi}_{46}+\frac{3}{4} \tilde{D}^{\phi}_{34} \tilde{D}^{\phi}_{22}-\Big(\frac{1}{2}\tilde{D}^{\phi}_{22}\Big)^3\Big]\beta_0\label{K46-bis}\\
 K_{45}&=&\Big[-\frac{3}{2}\tilde{D}^{\phi}_{45}+\frac{3}{4} \tilde{D}^{\phi}_{34} \tilde{D}^{\phi}_{21}+\frac{3}{4} \tilde{D}^{\phi}_{33} \tilde{D}^{\phi}_{22}-\frac{3}{2} \tilde{D}^{\phi}_{21}\Big(\frac{1}{2} \tilde{D}^{\phi}_{22}\Big)^2\Big]\beta_0\label{K45-bis}\\
 K_{44}&=&K^{ns}_{44}+K^{\phi, ns}_{44}+\Big[-\frac{3}{2}\tilde{D}^{\phi}_{44}+\frac{3}{4} \tilde{D}^{\phi}_{34} \tilde{D}^{\phi}_{20}+\frac{3}{4} \tilde{D}^{\phi}_{33} \tilde{D}^{\phi}_{21}+\frac{3}{4} \tilde{D}^{\phi}_{32} \tilde{D}^{\phi}_{22}\nonumber\\
 & &-\frac{3}{2} \tilde{D}^{\phi}_{22}\Big(\frac{1}{2} \tilde{D}^{\phi}_{21}\Big)^2-\frac{3}{2} \tilde{D}^{\phi}_{20}\Big(\frac{1}{2} \tilde{D}^{\phi}_{22}\Big)^2\Big]\beta_0+\Big[-\tilde{D}^{\phi}_{34}+\Big(\frac{1}{2}\tilde{D}^{\phi}_{22}\Big)^2\Big]\beta_1\label{K44-bis} \ ,\end{eqnarray}
 with similar, but longer relations (not written down for brevity) for $K_{4i}$ ($i\leq 3$).

\noindent If one now takes into account the known \cite{Soar:2009yh,Vogt:2010cv,Almasy:2010wn} single logarithmic behavior of $K_{2\phi}^{(i)}(N)$, i.e. the fact that $D^{\phi}_{ij}=0$ for $j\geq i$, one gets:

\begin{equation}K_{34}=K_{45}=K_{46}=0\label{K-null}\ ,\end{equation}
and:

\begin{equation}K_{ii}= K^{ns}_{ii}+K^{\phi, ns}_{ii}\label{Kii}\ .\end{equation}
Using the relations \cite{Soar:2009yh} $K^{ns}_{ii}=-4C_F\beta_0^{i-1}/i$ and $K^{\phi, ns}_{ii}=-4C_A\beta_0^{i-1}/i$, which follow (at the leading logarithmic level) from the   threshold resummation of the diagonal kernels  mentioned in Appendix A.1, one then recovers the results following from the assumed leading logarithmic threshold resummation of the singlet kernel $K$ (section 4.2). These results show that the leading logarithmic threshold resummation of $K$ follows from the single logarithmic enhancement of the off-diagonal physical kernel $K_{2\phi}$, together with the (leading logarithmic) threshold resummation of the diagonal kernels $K_{22}$ and $K_{\phi\phi}$. I note that eqs.(\ref{K11-bis})-(\ref{K30-bis})
allow also  to compute the $K_{ij}$'s, and thus check the results of section 3, using the results  for the $D^{\phi}_{i j}$'s  in \cite{Soar:2009yh}.

\noindent Quite similarly, one can show that the leading logarithmic threshold resummation of $K_{\phi}$ follows from the single logarithmic enhancement of the off-diagonal physical kernel $K_{\phi 2}$ (together with the leading logarithmic threshold resummation of the diagonal kernels $K_{22}$ and $K_{\phi\phi}$). The (now confirmed \cite{Vogt:2010cv,Almasy:2010wn}) ensuing  predictions  (eqs.(\ref{D46predict})-(\ref{D44predict})) of the double logarithmic off-diagonal four loop coefficients $D_{4i}$ ($4\leq i\leq 6$), as well as the similar predictions
(eqs.(\ref{Delta46predict})-(\ref{Delta44predict})) of the off-diagonal four loop coefficients  $\Delta_{4i}$ ($4\leq i\leq 6$)
 are thus equivalent to those of \cite{Soar:2009yh}.

\end{document}